\documentclass[a4, 11pt]{article} % amsart

\usepackage[utf8]{inputenc}
\usepackage[T1]{fontenc}
\usepackage{enumerate}
\usepackage{latexsym}
\usepackage[shortlabels]{enumitem}
\usepackage{graphicx}
\usepackage[colorlinks=true, urlcolor=blue, citecolor=blue, linkcolor=blue, raiselinks=true, breaklinks=true]{hyperref}
\usepackage[version=4]{mhchem}
\usepackage{siunitx}
\usepackage{bigints}
\usepackage{longtable,tabularx}

\usepackage{amssymb, amsmath, amsfonts, mathrsfs, amsthm, mathptmx}
\usepackage{esint}
\usepackage{cite}
\usepackage{mathtools}
\usepackage{subfig, placeins}
\usepackage{geometry}
\usepackage{tikz, tikz-3dplot}
\usetikzlibrary{patterns}
\usetikzlibrary{arrows}
\usepackage{authblk}
\usepackage[numbers, sort&compress]{natbib}

\usepackage{background}
\backgroundsetup{contents=Preprint}

\newcommand\norm[1]{\left\lVert#1\right\rVert}

\newtheorem{myTheorem}{Theorem}
\numberwithin{myTheorem}{section} % important bit

\newtheorem{myLemma}{Lemma}
\numberwithin{myLemma}{section} % important bit

\newtheorem{myRemark}{Remark}
\numberwithin{myRemark}{section} % important bit

\DeclareSymbolFont{newfont}{OML}{cmm}{m}{it}% Computer Modern math font
\DeclareMathSymbol{\varrho}{3}{newfont}{37}% Symbol 37

\newcommand*\diff{\mathop{}\!\mathrm{d}}

\newcommand{\Rint}{\mathop{\mathrlap{\pushpv}}\!\int}
\newcommand{\pushpv}{\mathchoice
  {\mkern5mu\rule[.6ex]{.5em}{1pt}}
  {\mkern2.8mu\rule[.5ex]{.35em}{.8pt}}
  {\mkern2.5mu\rule[.29ex]{.3em}{.7pt}}
  {\mkern2mu\rule[.2ex]{.2em}{.5pt}}}

\DeclareMathOperator{\Lin}{Lin}
\DeclareMathOperator{\Sym}{Sym}
\DeclareMathOperator{\Skw}{Skw}
\DeclareMathOperator{\tr}{tr}
\DeclareMathOperator{\diag}{diag}
\DeclareMathOperator{\meas}{meas}
\def\weak{\rightharpoonup}
\def\e{\varepsilon}

\begin{document}
	\sloppy
	
	\title{Linear Models of a Stiffened Plate via $\Gamma$-convergence}
		\author[1]{Marco Picchi Scardaoni\thanks{marco.picchiscardaoni@ing.unipi.it}}
	\author[1]{Roberto Paroni\thanks{roberto.paroni@unipi.it}}
	\affil[1]{Università di Pisa, Dipartimento di Ingegneria Civile e Industriale, Largo Lucio Lazzarino 2, 56122, Pisa}
	
	\date{\today}
	\maketitle

	\begin{abstract}
		We consider a family of three-dimensional stiffened plates whose dimensions are scaled through different powers of a small parameter $\e$. The plate and the stiffener  are assumed to be linearly elastic, isotropic, and homogeneous. By means of $\Gamma$-convergence, we study the asymptotic behavior of the three-dimensional problems as the parameter $\e$ tends to zero. 
For different relative values of the powers of the parameter $\e$, we show how the interplay between the plate and the stiffener affects the limit energy. We derive twenty-three limit problems.

		
	\end{abstract}
	
	\section{Introduction} \label{intro}
	Thin-walled structures are widely used in many engineering fields, such as aeronautic and aerospace structures, vessels, civil and mechanical constructions.
	Since the widespread use of such structures, many models, based on \textit{a priori} kinematical assumptions, have been proposed in the history of Mechanics, in order to predict the behavior of loaded structures.
	Even though these models have been used successfully by generations of practitioners, they generally rely on heuristic assumptions. From a theoretical point of view, as reasonable as these assumptions may sound, they are hypotheses that jeopardize the validity of the mechanical models.
	
	Over the last few years, attention has been paid to the rigorous justification of the classical mechanical theories and models: beams, shells, plates, etc. The underlying idea is to study the asymptotic behavior of \textit{actual} three-dimensional variational formulations and to let some "smallness parameter" go to zero, so to fetch the essential features of the primitive problem in the resulting "simplified" asymptotic one.
	
	A way to justify the mechanical models is via $\Gamma$-convergence, a variational convergence notion appeared for the first time in a the seminal work by De Giorgi and Franzoni \cite{Degiorgi1975}. 
	The underlying idea is to replace the functional ruling the actual problem at hand by a new one, more handy, such that it may capture the major features of the primitive problem.
 More in detail, $\Gamma$-convergence derives the aimed asymptotic functional in such a way to achieve the convergence of minima and minimizers of the primitive problem and of the asymptotic one. The reader is addressed, for instance, to the monographs by Dal Maso and Braides \citep{Maso1993, braides} for an exhaustive exposition of the topic.\\
	Many interesting results have been obtained in Mechanics by $\Gamma$-convergence. 
	One of the first works are by Acerbi \textit{et al.} \cite{Acerbi} and by Anzellotti \textit{et al.} \cite{anzellotti}. In the former, the asymptotic behavior of a string is derived in the framework of non-linear elasticity. In the latter, it is studied a plate and a beam in the context of linearized elasticity. These works represent a milestone, since they developed a fundamental \textit{modus operandi} for the following works.
	$\Gamma$-convergence has been successfully used to justify models for beams and plates in linear and non-linear elasticity, in isotropic and anisotropic elasticity, also under residual stress condition \cite{Freddi2013, Freddi2004, Paroni2009, Paroni2007, ppg2, Davini2008, DalMaso2002, Friesecke2002, Mora2004}.
	By $\Gamma$-convergence, there have been justified also more complex theories, such as the Vlassov theory of thin-walled beams and the well-known Bredt formul\ae~for torsion \cite{Freddi2007}.
	
	In model dimension reduction, a special chapter is reserved to junctions between bodies, possibly having different asymptotic dimensions: for instance, the asymptotic study of the junction of a 3D-body with a 2D-body. Several contributes can be counted \cite{ledret, Ledret1989, ciarletjunction, Ciarlet1989, Gruais1993, Gruais1993a}. More recently, \cite{Gaudiello2007, Nardinocchi2002, Leugering2018, Gaudiello2010} have studied the junction between multi-domain bodies using classical variational techniques.\\
	Remarkably, among the works about the limit models of joined plates and beams, there is a lack of asymptotic models of stiffened plates. At the best of the authors' knowledge, there is one work in the literature facing this problem of practical interest \cite{aufranc}. However, in \cite{aufranc}, the author considers only a special case, since the beam cross-section and the plate thickness scale with the same order of magnitude. Moreover, the limit behavior is studied without taking into account  the torsion angle of the stiffener, and over the junction region the plate and the stiffener assume different elastic moduli, which is physically implausible.\\
	Hereby, we face the problem of deriving the asymptotic model of a stiffened plate in the framework of linear isotropic elasticity, as the plate thickness and the beam cross-section go to zero, possibly with different scaling velocities. Moreover, the torsion angle limit derivation is deeply discussed. The variational convergence is obtained in the sense of $\Gamma$-convergence.
		
	This paper is organized as follows. Section \ref{sec:notation} introduces the  notation adopted in this work, together with the principal functional spaces. Section \ref{sec:setting} introduces the setting of the variational problem, whilst Section \ref{sec:scaledproblem} introduces the mapping from the three-dimensional structure to a family of thin domains together with two compactness results. Section  \ref{sec:junction} is dedicated to the limit joining conditions of displacements and stiffener cross-sectional rotation angle. In Section \ref{sec:limitenergy} the expression of the limit energy is obtained and the main $\Gamma$-convergence Theorem is proved. Finally, Section \ref{sec:congergence}  shows that the convergence of minima and minimizers is actually strong.

\section{Notation}\label{sec:notation}
In this paper, we work in the real Euclidean three-dimensional space $\mathbb{R}^3$.
We use upper-case bold letters to indicate tensors and lower-case bold letters to indicate vectors. The Euclidean (Frobenius) product is indicated with $\cdot$ and the corresponding induced norm by $|\cdot|$. We denote by $\Lin$ the set of linear transformations, and by $\Sym$, $\Skw$ the subsets of linear symmetric and antisymmetric ones, respectively.  $\tr(\cdot)$ denotes the trace operator, whilst $\diag(a, b, c)$ is the diagonal matrix with elements $a$, $b$, $c$ on the principal diagonal. $\mathbb{R}^+$ denotes the set of all strictly positive real numbers, while $\mathbb{N}$ denotes the non-negative integer ones. \\
Let $S \subset \mathbb{R}^n$ ($n \in \{1, 2, 3\}$) be open. For any function $\mathbf{v}:S \to \mathbb{R}^3$, we shall denote its gradient by
\begin{equation*}
	\mathbf{H}\mathbf{v}\coloneqq\nabla \mathbf{v},
\end{equation*}
and its unique decomposition in a symmetric and skew-symmetric part by
$
\mathbf{H}\mathbf{v} = \mathbf{E}\mathbf{v} + \mathbf{W}\mathbf{v},
$
where
\begin{equation*}
	\begin{aligned}
		\mathbf{E}\mathbf{v}&\coloneqq\frac{\nabla \mathbf{v}+\nabla \mathbf{v}^T}{2},
		&\mathbf{W}\mathbf{v}&\coloneqq\frac{\nabla \mathbf{v}-\nabla \mathbf{v}^T}{2}.
	\end{aligned}
\end{equation*}

We denote by
\begin{equation*}
	L^2(S, \mathbb{R}^q)\coloneqq\left\lbrace \mathbf{v}: S \to \mathbb{R}^q\,:\,\norm{\mathbf{v}}_{L^2(S, \mathbb{R}^q)} < \infty \right\rbrace 
\end{equation*}
the Banach space of (equivalent classes of) Lebesgue-integrable functions on $S$ with values in $\mathbb{R}^q$ ($q\in \mathbb{N}\setminus  \{0\}$), where $$\norm{\mathbf{v}}_{L^2(S, \mathbb{R}^q)}\coloneqq \left(\int_S |\mathbf{v}|^2\right)^{1/2}.$$\\
The corresponding Sobolev' spaces of functions on $S$ with values in $\mathbb{R}^q$ are the Banach spaces defined as follows (we will need only the cases for which $l\in \{1, 2\}$): 
\begin{equation*}
	W^{l,2}(S, \mathbb{R}^q)\coloneqq\left\lbrace \mathbf{v}: S \to \mathbb{R}^q\,:\, \mathbf{v}\in L^2(S, \mathbb{R}^q), \nabla^\alpha \mathbf{v} \in L^2(S, \mathbb{R}^{n^\alpha\times q}) \forall \alpha(\in\mathbb{N}) \leq l \right\rbrace.
\end{equation*} 
They are endowed with the norm $$\norm{\mathbf{v}}^2_{W^{l,2}(S, \mathbb{R}^q)}\coloneqq \norm{\mathbf{v}}^2_{L^2(S, \mathbb{R}^q)} + \sum\limits_{\alpha=1}^l \norm{\nabla^\alpha\mathbf{v}}^2_{L^2(S, \mathbb{R}^{n^\alpha\times q})}.$$\\
Note that the (high-order) gradients $\nabla^\alpha(\cdot)$ shall be understood in the sense of distributions.\\
We shall furthermore consider the Sobolev' spaces  
\begin{equation*}
	W^{l,2}_{0}(S, \mathbb{R}^q)\coloneqq\{\mathbf{v}\in W^{l,2}(S,\mathbb{R}^q):\mathbf{v}=\mathbf{0} \textrm{ in } \partial_D S\}
\end{equation*}
as the set of functions belonging to $W^{l,2}(S, \mathbb{R}^q)$ that assume value zero on a certain subset $\partial_D S$ of the boundary of $S$. 
In this paper, we will make use of standard results concerning Sobolev' spaces. The reader is addressed to the classical monograph by Adams \cite{Adams2003} or to the more recent book by Leoni \cite{Leoni2017}. 

If $r\in \mathbb{N}\cup \{\infty\}$, then $C^r(S, \mathbb{R}^q)$ denotes the space of $r$-times continuously differentiable functions on $S$ with values in $\mathbb{R}^q$, and $C^\infty_0(S, \mathbb{R}^q)$ denotes the space of functions belonging to $C^\infty(S, \mathbb{R}^q)$ that assume value zero in a neighborhood of the boundary of $S$.\\
We will refrain to specify the codomain $\mathbb{R}$ in the notation of the functional spaces: for instance, we will simply write $W^{1,2}(S)$ instead of $W^{1,2}(S, \mathbb{R})$, and so forth.

We will denote by $W^{-1,2}$ the dual space of $W^{1,2}_0$. We recall that the following compact embeddings hold $C_0^\infty \hookrightarrow W^{1,2}_0 \hookrightarrow L^2 \hookrightarrow W^{-1,2}$, and that, if the operator $T\in W^{-1,2}$,
\begin{equation*}
	\| T \|_{W^{-1,2}} \coloneqq \sup\limits_{x\in W^{1,2}_0, \, x\neq 0} \frac{\left|<T,x>\right|}{\| x\|_{W^{1,2}_0}},
\end{equation*}
where $<T,x>$ denotes the dual pairing $W^{-1,2}\times W^{1,2}_0$.

If not specified, we adopt Einstein' summation convention for indices. Indices $\alpha,\beta,\gamma,\delta$ take values in the set $\{1,2\}$, indices $a, b, c, d$ in the set $\{2,3\}$, and indices $i,j$ in the set $\{1,2,3\}$.
With the notation $A\lesssim B$ we mean that there exists a constant $C>0$ such that $A \leq C\,B$. Such constant may vary line to line. 
As it is usual, we denote by  $\Rint_{S} f(\cdot) \diff x$ the average value of the  function $f(\cdot)$ over its integration domain, i.e. $\frac{1}{\meas(S)}\int_S f(\cdot) \diff x$, $\meas(S)$ being the Lebesgue measure of the set $S$. We denote the strong convergence (convergence in norm) with the symbol $\to$, whilst the weak convergence will be denoted by $\weak$. Finally, with the symbol $x\downarrow y$, we mean that $x$ is approaching $y$ from above (i.e., $x\to y^+$).

\section{General Setting}\label{sec:setting}
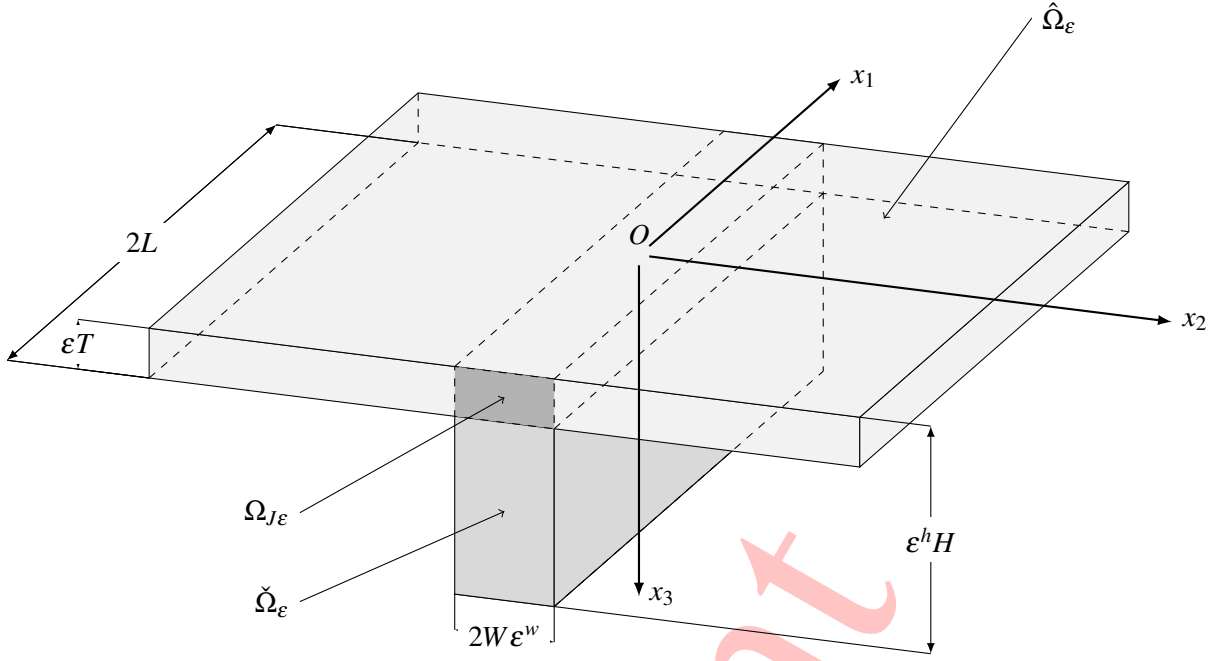
\begin{figure}[!hbt]
	\centering
		\tikzset{
		load/.style   = {ultra thick,-latex},
		stress/.style = {-latex},
		dim/.style    = {latex-latex},
		axis/.style   = {-latex,black!55},
		    dimen/.style={<->,>=latex,thin,every rectangle node/.style={fill=white,midway,font=\sffamily}},
	}
	\tikzstyle{dimetric2}=[x={(+0.354cm,+0.312cm)},z={(0cm,-0.943cm)},y={(0.935cm,-0.118cm)}]
	\def\T{0.7}
	\def\L{5}
	\def\W{0.7}
	\def\H{2.5}
	\begin{tikzpicture}[dimetric2]
	\node (origin) at (0,0,0) {}; % shift relative baseline

	\coordinate (A) at (-\L,\W,\T+\H);
	\coordinate (B) at (\L,\W,\T+\H);
	\coordinate (C) at (\L,-\W,\T+\H);
	\coordinate (D) at (-\L,-\W,\T+\H);
	\coordinate (E) at (-\L,\W,0);
	\coordinate (F) at (\L,\W,0);
	\coordinate (G) at (\L,-\W,0);
	\coordinate (H) at (-\L,-\W,0);
	\coordinate (I) at (-\L,\L,\T);
	\coordinate (L) at (\L,\L,\T);
	\coordinate (M) at (\L,-\L,\T);
	\coordinate (N) at (-\L,-\L,\T);
	\coordinate (O) at (-\L,\L,0);
	\coordinate (P) at (\L,\L,0);
	\coordinate (Q) at (\L,-\L,0);
	\coordinate (R) at (-\L,-\L,0);

	% TRAVE
	\draw[fill=gray!30]  (A)--(B)--(C)--(D)--cycle;
	\draw[fill=gray!30]  (A)--(E)--(H)--(D)--cycle;
	\draw[fill=gray!30]  (B)--(F)--(G)--(C)--cycle;
	\draw[fill=gray!30]  (A)--(B)--(F)--(E)--cycle;
	
	%piastra
	\draw[fill=gray!10]  (I)--(O)--(R)--(N)--cycle;
	\draw[fill=gray!10]  (I)--(L)--(P)--(O)--cycle;
	\draw[fill=gray!10]  (O)--(P)--(Q)--(R)--cycle;

	% assi e tratteggio
	\draw[axis,black,thick] (origin) -- ++(1.5*\L,0,0) node[right] {$x_1$};
	\draw[axis,black,thick] (origin) -- ++(0,1.5*\L,0) node[right] {$x_2$};
	\draw[dashed] (E) -- (F);
	\draw[dashed] (H) -- (G);
	\draw[dashed] (Q) -- (M);
	\draw[dashed] (N) -- (M);
	\draw[dashed] (L) -- (M);
	\draw[dashed] (A) -- (B);
	\draw[dashed] (F) -- (B);
	
	\draw[dashed] (-\L, \W, \T) -- (\L, \W, \T);
	
	\draw[axis,black, thick] (origin) -- ++(0,0,1.5*\H+1.5*\T) node[right] {$x_3$};

	\draw[thin] (A) -- (-\L,\W,\T+\H+0.5);
	\draw[thin] (D) -- (-\L,-\W,\T+\H+0.5);
	\draw[dimen] (-\L,-\W,\T+\H+0.5) -- (-\L,\W,\T+\H+0.5) node{$2W\e^w$};
	
	\draw[thin] (N) -- (-\L,-\L-2,\T);
	\draw[thin] (M) -- (\L,-\L-2,\T);
	\draw[dimen] (-\L,-\L-2,\T) -- (\L,-\L-2,\T) node{$2L$};

	\draw[thin] (N) -- (-\L,-\L-1,\T);
	\draw[thin] (R) -- (-\L,-\L-1,0);
	\draw[dimen] (-\L,-\L-1,\T) -- (-\L,-\L-1,0) node{$\e T$};
	
	\draw[thin] (N) -- (-\L,-\L-2,\T);
	\draw[thin] (M) -- (\L,-\L-2,\T);
	\draw[dimen] (-\L,-\L-2,\T) -- (\L,-\L-2,\T) node{$2L$};
	
	\draw[thin] (O) -- (-\L,\L+1,0);
	\draw[thin] (A) -- (-\L,\L+1,\H+\T);
	\draw[dimen] (-\L,\L+1,0) -- (-\L,\L+1,\H+\T) node{$\e^h H$};

	\draw[fill=gray!60, dashed]  (E)--(H)--(-\L, -\W, \T)--(-\L, \W, \T)--cycle;
		
	\coordinate[label=$O$] (origin);

		\draw[->] (\L+3,\L/2,-1) -- (\L/2,\L/2,0) node[right] at (\L+3,\L/2,-1) {$\hat{\Omega}_\e$};
		\draw[->] (-\L-1,-\L/2,+\T+\H) -- (-\L,0,+\T+\H/2) node[left] at (-\L-1,-\L/2,+\T+\H) {$\check{\Omega}_\e$};
		\draw[->] (-\L-1,-\L/2,+\T+\H/2) -- (-\L,0,\T/2) node[left] at (-\L-1,-\L/2,+\T+\H/2) {$\Omega_{J\e}$};
	\end{tikzpicture}
	\caption{Geometry of the real problem}
	\label{fig:piastra}
\end{figure}

Let us introduce the real parameter $\e$ that takes values in a sequence of positive numbers converging to zero. 
With reference to Fig.~\ref{fig:piastra}, we introduce in $\mathbb{R}^3$ an orthonormal absolute reference system, denoted by $(O; x_1, x_2, x_3)$. We consider a plate-like  body (hereafter, with a slightly abuse of language, just \textit{plate}) occupying the region $\hat{\Omega}_\e \coloneqq(-L, L)\times(-L,L)\times \e(0, T)$ and a blade-like stiffener body (hereafter, with a slightly abuse of language, just \textit{stiffener}) occupying the region $\check{\Omega}_\e \coloneqq (-L,L)\times \e^w(-W , W)\times \e^h(0, H)$, being $L, T, H, w,\,h \in \mathbb{R}^+$. Moreover, let $\Omega_{J\e}\coloneqq \hat{\Omega}_\e \cap \check{\Omega}_\e = (-L,L)\times \e^w(-W, W)\times \e(0, T)$ be the overlapping region, that hereafter we refer to as \textit{junction region}.\\
We assume that $h<1$ and that $W<L$. The first assumption implies that the height of the stiffener is larger than the thickness of the plate, while the second assumption is simply made to assure that the plate is larger than the width, $\e^w W$, of the stiffener even in the case $w=0$.  The  domain $\Omega_\e \coloneqq \hat{\Omega}_\e \cup \check{\Omega}_\e$ is depicted in Fig.~\ref{fig:piastra}. We shall consider the body clamped at $x_1=L$, i.e., the displacement field is null in all points with coordinate $x_1$ equal to $L$ (hence $\partial_D S \coloneqq  \Omega_\e \cap \{x_1=L\}$): this condition will be hereafter referred to as \textit{boundary condition}.\\

In what follows, we denote the cross-section of the stiffener by
$
\check{\omega}_\e \coloneqq \e^w(-W , W)\times \e^h(0, H)
$ and the mid-plane of the plate as $
\hat{\omega} \coloneqq (-L,L)\times (-L , L)
$.

The stored-energy functional $\widetilde{\mathcal{W}}_\e:\Omega_\e\to\mathbb{R}^+$ is defined by
\begin{equation}\label{eq:firstenergy}
\widetilde{\mathcal{W}}_\e(\mathbf{v})\coloneqq\frac{1}{2}\int_{\Omega_\e} \mathbb{C}\left[\mathbf{Ev}\right]\cdot \mathbf{Ev}\, \diff x,
\end{equation}
where $\mathbb{C}$ is a fourth-order elasticity tensor, positive definite and having the usual major and minor symmetries. In particular, this implies that there exists a positive constant $ \mu$ such that
\begin{equation}\label{coercivity}
	\begin{aligned}
\mathbb{C}[\mathbf{A}]\cdot \mathbf{A} &\geq \mu |\mathbf{A}|^2, &\forall \mathbf{A} \in \Sym.
\end{aligned}
\end{equation} 
In this paper, we consider a linear homogeneous isotropic material. It can be shown (see, for instance, \cite[Article 68]{Love1944}) that for this kind of materials the stored energy density admits the unique representation 
\begin{equation}\label{f}
		\begin{aligned}
		f(\mathbf{A}) &\coloneqq \frac{1}{2}\mathbb{C}[\mathbf{A}]\cdot\mathbf{A}= \mu |\mathbf{A}|^2 +\frac{\lambda}{2} |\tr\mathbf{A}|^2, &\forall \mathbf{A} \in \Sym,
	\end{aligned}
\end{equation}
where $\mu>0$ and $\lambda>-\frac{2}{3}\mu$ are called Lamé parameters. The energy density \eqref{f} is also known as Saint Venant-Kirchhoff's.

The functional in \eqref{eq:firstenergy} can be decomposed into the sum of two contributions
\begin{equation*}
	\widetilde{\mathcal{W}}_\e(\mathbf{v}) = \frac{1}{2}\int_{\hat{\Omega}_\e} \chi_\e(x) \mathbb{C}\left[\mathbf{Ev}\right]\cdot \mathbf{Ev}\, \diff x + \frac{1}{2}\int_{\check{\Omega}_\e} \chi_\e(x) \mathbb{C}\left[\mathbf{Ev}\right]\cdot \mathbf{Ev}\, \diff x,
\end{equation*}
where $\chi_\e(x) : \Omega_\e \to\{\frac{1}{2},1\}$ is defined by
\begin{equation*}
	\chi_\e(x) \coloneqq \begin{cases}
		\frac{1}{2}, & \mathrm{ if } \ x \in \Omega_{J\e}, \\
		1, &\text{otherwise}.
	\end{cases}
\end{equation*}
The function $\chi_\e$ allows to associate half of the energy of the junction region to the stiffener and the other half to the plate, or, more simply, it avoids to consider the energy of the region $\Omega_{J\e}$ twice.\\

\section{The Scaled Problem}\label{sec:scaledproblem}

As it is usual in this kind of problems, we change variables and define the energies over  fixed domains (see \cite{Ciarlet1979, Acerbi}). For this purpose, we set
$$
\hat{\Omega}\coloneqq\hat{\Omega}_1, \quad\check{\Omega}\coloneqq\check{\Omega}_1, \quad\Omega\coloneqq\Omega_1,\quad\Omega_{J} \coloneqq\Omega_{J 1},\quad\check{\omega} \coloneqq \check{\omega}_1.
$$

For the plate, we introduce the scaling mapping $\hat{\mathbf{q}}_\e :\hat{\Omega}\to\hat{\Omega}_\e$ defined by $\hat{\mathbf{q}}_\e(x_1,x_2,x_3)\coloneqq(x_1,x_2,\e x_3)$. After setting $\hat{\mathbf{Q}}_\e \coloneqq \nabla\hat{\mathbf{q}}_\e = \diag(1,1,\e)$, we define $\hat{\mathbf{u}}_\e :\Omega \to \mathbb{R}^3$ by
\begin{equation}\label{eq:scalingpiastra}
\begin{split}
\hat{\mathbf{u}}_\e &\coloneqq \hat{\mathbf{Q}}_\e \mathbf{u} \circ \hat{\mathbf{q}}_\e,
\end{split}
\end{equation}
for every function $\mathbf{u}:\Omega_\e \to \mathbb{R}^3$.
Taking derivatives, one finds the scaled gradient for the plate
\begin{equation}\label{Hhat}
\hat{\mathbf{H}}_\e \hat{\mathbf{u}}_\e  \coloneqq \left(\mathbf{H}\mathbf{u}\right)\circ \hat{\mathbf{q}}_\e = \hat{\mathbf{Q}}_\e^{-1} \mathbf{H} \hat{\mathbf{u}}_\e  \hat{\mathbf{Q}}_\e^{-1} =
\left(\begin{matrix}
(\mathbf{H}\hat{\mathbf{u}}_\e)_{11}& (\mathbf{H}\hat{\mathbf{u}}_\e)_{12} & \frac{(\mathbf{H}\hat{\mathbf{u}}_\e)_{13}}{\e}\\[10pt]
(\mathbf{H}\hat{\mathbf{u}}_\e)_{21}& (\mathbf{H}\hat{\mathbf{u}}_\e)_{22}& \frac{(\mathbf{H}\hat{\mathbf{u}}_\e)_{23}}{\e}\\[10pt]
\frac{(\mathbf{H}\hat{\mathbf{u}}_\e)_{31}}{\e} &
\frac{(\mathbf{H}\hat{\mathbf{u}}_\e)_{32}}{\e}  & \frac{(\mathbf{H}\hat{\mathbf{u}}_\e)_{33}}{\e^2}\\
\end{matrix}\right),
\end{equation}
and the scaled strain for the plate
\begin{equation}\label{eq:strainpiastra}
\hat{\mathbf{E}}_\e \hat{\mathbf{u}}_\e  \coloneqq \left(\mathbf{E}\mathbf{u}\right)\circ \hat{\mathbf{q}}_\e = \hat{\mathbf{Q}}_\e^{-1} \mathbf{E} \hat{\mathbf{u}}_\e  \hat{\mathbf{Q}}_\e^{-1}.
\end{equation}

Similarly, for the stiffener, we introduce the scaling mapping
$\check{\mathbf{q}}_\e:\check{\Omega}\to\check{\Omega}_\e$ defined by $\check{\mathbf{q}}_\e(x_1,x_2,x_3)\coloneqq(x_1,\e^w x_2,\e^h x_3)$, the tensor $\check{\mathbf{Q}}_\e \coloneqq \nabla\check{\mathbf{q}}_\e = \diag(1,\e^w, \e^h)$, and the scaled displacement $\check{\mathbf{u}}_\e: \Omega\to \mathbb{R}^3$ defined as
\begin{equation}\label{eq:scalingtrave}
\check{\mathbf{u}}_\e \coloneqq \check{\mathbf{Q}}_\e  \mathbf{u} \circ \check{\mathbf{q}}_\e.
\end{equation}
The scaled gradient for the stiffener is defined by
\begin{equation}\label{eq:Hcheck}
\check{\mathbf{H}}_\e \check{\mathbf{u}}_\e \coloneqq \left(\mathbf{H}\mathbf{u}\right)\circ \check{\mathbf{q}}_\e = \check{\mathbf{Q}}_\e^{-1} \mathbf{H} \check{\mathbf{u}}_\e \check{\mathbf{Q}}_\e^{-1} =
\left(\begin{matrix}
(\mathbf{H} \check{\mathbf{u}}_\e)_{11}& 
\frac{(\mathbf{H} \check{\mathbf{u}}_\e)_{12} }{\e^w} & \frac{(\mathbf{H} \check{\mathbf{u}}_\e)_{13}}{\e^h}\\[10pt]
\frac{(\mathbf{H} \check{\mathbf{u}}_\e)_{21} }{\e^w}& \frac{(\mathbf{H} \check{\mathbf{u}}_\e)_{22} }{\e^{2w}}& \frac{(\mathbf{H} \check{\mathbf{u}}_\e)_{23}}{\e^{w+h}}\\[10pt]
\frac{(\mathbf{H} \check{\mathbf{u}}_\e)_{31}}{\e^h} &
\frac{(\mathbf{H} \check{\mathbf{u}}_\e)_{32}}{\e^{w+h}}  & \frac{(\mathbf{H} \check{\mathbf{u}}_\e)_{33}}{\e^{2h}}\\
\end{matrix}\right),
\end{equation}
so that the scaled strain for the stiffener reads
\begin{equation}\label{eq:straintrave}
\check{\mathbf{E}}_\e \check{\mathbf{u}}_\e \coloneqq \left(\mathbf{E}\mathbf{u}\right)\circ \check{\mathbf{q}}_\e = \check{\mathbf{Q}}_\e^{-1} \mathbf{E} \check{\mathbf{u}}_\e \check{\mathbf{Q}}_\e^{-1}.
\end{equation}

By changing variables, we rewrite the stored energy over the fixed domains, dividing by $\e$ that does not affect what follows:
\begin{equation}\label{energy}
\begin{split}
 \mathcal{W}_\e(\hat{\mathbf{u}}_\e,\check{\mathbf{u}}_\e) \coloneqq  \frac{\widetilde{\mathcal{W}}_\e}{\e}  &=\frac{1}{2}\int_{\hat{\Omega}}\hat{\chi}_\e\mathbb{C}\left[\hat{\mathbf{E}}_\e\hat{\mathbf{u}}_\e\right]\cdot \hat{\mathbf{E}}_\e\hat{\mathbf{u}}_\e \diff x  +\frac{1}{2}\int_{\check{\Omega}} \check{\chi}_\e\mathbb{C}\left[\e^{k}\check{\mathbf{E}}_\e\check{\mathbf{u}}_\e\right]\cdot \e^{k}\check{\mathbf{E}}_\e\check{\mathbf{u}}_\e \diff x,\\
&=:\hat{\mathcal{W}}_\e(\hat{\mathbf{u}}_\e)+\check{\mathcal{W}}_\e(\e^k\check{\mathbf{u}}_\e),
\end{split}
\end{equation}
where we have set $ \hat{\chi}_\e \coloneqq \chi_\e \circ \hat{\mathbf{q}}_\e$, $ \check{\chi}_\e \coloneqq \chi_\e \circ \check{\mathbf{q}}_\e$ and 
\begin{equation*}
	2k\coloneqq w+h-1.
\end{equation*}

We now prove the compactness of some properly rescaled sequences of displacements, both for the plate and the stiffener.
The limit displacement fields will be of Kirchhoff-Love and Bernoulli-Navier type. The former is defined by
\begin{equation*}
KL_0(\hat{\Omega}_\e)\coloneqq\left\lbrace \mathbf{v}\in W^{1,2}_0(\hat{\Omega}_\e, \mathbb{R}^3):\exists\, \hat{\xi}_\alpha\in W^{1,2}_0(\hat{\omega}), \exists \ \hat{\xi}_3\in W^{2,2}_0(\hat{\omega}), v_\alpha=\hat{\xi}_\alpha - x_3 \partial_\alpha\hat{\xi}_{3},  v_3=\hat{\xi}_3 \right\rbrace,
\end{equation*}  
the latter by
\begin{equation*}
BN_0(\check{\Omega}_\e)\coloneqq\left\lbrace \mathbf{v}\in W^{1,2}_0(\check{\Omega}_\e, \mathbb{R}^3): \exists \ \check{\xi}_1\in W^{1,2}_0((-L,L)), \exists\, \ \check{\xi}_a \in W^{2,2}_0((-L,L)), v_1= \check{\xi}_1 -x_2\partial_1\check{\xi}_{2}-x_3\partial_1\check{\xi}_{3}, \ v_a=\check{\xi}_a\right\rbrace .
\end{equation*}

\begin{myLemma}\label{compattezzapiastra}
Let $\{\hat{\mathbf{u}}_\e\} \subset W^{1,2}_0 (\hat{\Omega}, \mathbb{R}^3) $ be a sequence such that $\sup_\e \hat{\mathcal{W}}_\e(\hat{\mathbf{u}}_\e) < \infty$. Then, there exist a subsequence (not relabeled) $\{\hat{\mathbf{u}}_{\e}\} $ and a $\hat{\mathbf{u}}\in \text{KL}_0(\hat{\Omega})$ such that $\hat{\mathbf{u}}_{\e} \weak \hat{\mathbf{u}}$ in $W^{1,2}(\hat{\Omega},\mathbb{R}^3)$. We denote by $\hat{\mathbf{E}}$ the limit of $\hat{\mathbf{E}}_\e \hat{\mathbf{u}}_\e$ in the weak topology of  $L^2(\hat{\Omega}, \mathbb{R}^{3\times 3})$.
\end{myLemma}
\begin{proof}
Since $\mathbb{C}$ is positive-definite, from inequality \eqref{coercivity} and from Korn inequality, it follows that
\begin{equation*}
\sup_\e \norm{ \hat{\mathbf{u}}_\e}_{W^{1,2}(\hat{\Omega},\mathbb{R}^3)}^2 \lesssim    \sup_\e \norm{\mathbf{E}\hat{\mathbf{u}}_\e}_{L^2(\hat{\Omega}, \mathbb{R}^{3\times 3})}^2 
\lesssim    \sup_\e \norm{\hat{\mathbf{E}}_\e \hat{\mathbf{u}}_\e}_{L^2(\hat{\Omega}, \mathbb{R}^{3\times 3})}^2< \infty.
\end{equation*}
Hence, up to a subsequence, $\hat{\mathbf{u}}_{\e} \weak \hat{\mathbf{u}}$ for a certain $\hat{\mathbf{u}}\in W^{1,2}_0(\hat{\Omega}, \mathbb{R}^{3})$. From \eqref{eq:strainpiastra}, we have that
$\norm{(\mathbf{E}\hat{\mathbf{u}}_\e)_{\alpha 3}}_{L^2(\hat{\Omega})} = \norm{\e (\hat{\mathbf{E}}_\e\hat{\mathbf{u}}_\e)_{\alpha 3}}_{L^2(\hat{\Omega})} \lesssim \e$ and 
$\norm{(\mathbf{E}\hat{\mathbf{u}}_\e)_{3 3}}_{L^2(\hat{\Omega})} = \norm{\e^2 (\hat{\mathbf{E}}_\e\hat{\mathbf{u}}_\e)_{3 3}}_{L^2(\hat{\Omega})} \lesssim \e^2$. As a consequence, we deduce that
\begin{equation*}
\begin{aligned}
\partial_3\hat{u}_{ \alpha\e} +\partial_\alpha \hat{u}_{3\e} &\to\partial_3\hat{u}_{\alpha} +\partial_\alpha\hat{u}_{3} = 0, && \textrm{ in } L^2(\hat{\Omega}) \ \textrm{ (not summed on $\alpha$)},\\
\partial_3\hat{u}_{3\e}  &\to \partial_3\hat{u}_{3} = 0, && \textrm{  in } L^2(\hat{\Omega}).
\end{aligned}
\end{equation*}
Hence, $\hat{\mathbf{u}}\in KL_0(\hat{\Omega})$ as follows by integration.
We conclude the lemma by noticing that $\hat{\mathbf{E}}_\e \hat{\mathbf{u}}_\e \weak \hat{\mathbf{E}}$ in $L^2(\hat{\Omega}, \mathbb{R}^{3\times 3})$ for some $\hat{\mathbf{E}}$ in $L^2(\hat{\Omega}, \mathbb{R}^{3\times 3})$.
%\QEDN
\end{proof}

\begin{myLemma}\label{compattezzatrave}
Let $\{ \e^k \check{\mathbf{u}}_\e\} \subset W^{1,2}_0 (\check{\Omega},\mathbb{R}^3) $ be a sequence such that $\sup_\e \check{\mathcal{W}}_\e(\e^k\check{\mathbf{u}}_\e) < \infty$; then, there exist a subsequence (not relabeled) $\{\e^k\check{\mathbf{u}}_{\e}\} $ and a $\check{\mathbf{u}}\in BN_0 (\check{\Omega})$ such that $\e^k\check{\mathbf{u}}_{\e} \weak \check{\mathbf{u}}$ in $W^{1,2}(\check{\Omega},\mathbb{R}^3)$. We denote by $\check{\mathbf{E}}$ the limit of $\e^k\check{\mathbf{E}}_\e \check{\mathbf{u}}_\e$ in the weak topology of  $L^2(\check{\Omega}, \mathbb{R}^{3\times 3})$.
\end{myLemma}
\begin{proof}
The proof is similar to that of Lemma \ref{compattezzapiastra}. Since $\mathbb{C}$ is positive-definite and from Korn inequality, it follows that
\begin{equation*}
 \sup_\e \norm{ \e^k\check{\mathbf{u}}_\e}_{W^{1,2}(\check{\Omega}, \mathbb{R}^{3})}^2 \lesssim    \sup_\e \norm{ \e^k\mathbf{E} \check{\mathbf{u}}_\e}_{L^2(\check{\Omega}, \mathbb{R}^{3\times 3})}^2 
 \lesssim    \sup_\e \norm{ \e^k\check{\mathbf{E}}_\e \check{\mathbf{u}}_\e}_{L^2(\check{\Omega}, \mathbb{R}^{3\times 3})}^2 < \infty.
\end{equation*}
Hence, up to a subsequence, $\e^k\check{\mathbf{u}}_{\e} \weak \check{\mathbf{u}}$ for a certain $\check{\mathbf{u}}\in W^{1,2}_0(\check{\Omega}, \mathbb{R}^{3})$. Since $\{\left(\e^k \mathbf{E} \check{\mathbf{u}}_\e\right)_{ij}\}$ for $(i,j)\neq (1,1)$ converge to zero in $L^2(\check{\Omega})$, we deduce that
\begin{equation*}
\begin{aligned}
\e^k\left(\partial_a\check{u}_{1\e} +\partial_1\check{u}_{a\e}\right) &\to \partial_a\check{u}_{1} +\partial_1\check{u}_{a} = 0, && \textrm{ in } L^2(\check{\Omega}) \ \textrm{(not summed on a)},\\
\e^k\left(\partial_2\check{u}_{3\e} +\partial_3\check{u}_{2\e}\right) &\to \partial_2\check{u}_{3} +\partial_3\check{u}_{2} = 0, && \textrm{ in } L^2(\check{\Omega}),\\
\e^k\partial_a\check{u}_{a\e} &\to \partial_a\check{u}_{a} = 0, && \textrm{ in } L^2(\check{\Omega}) \ \textrm{(not summed on a)}.
\end{aligned}
\end{equation*}
Hence,  $\check{\mathbf{u}}\in BN_0(\check{\Omega})$ as follows by integration. We conclude the lemma by noticing that $\e^k \check{\mathbf{E}}_\e \check{\mathbf{u}}_\e \weak \check{\mathbf{E}}$ in  $L^2(\check{\Omega}, \mathbb{R}^{3\times 3})$ for some $\check{\mathbf{E}}$ in $L^2(\check{\Omega}, \mathbb{R}^{3\times 3})$.
\end{proof}

\section{The Junction Conditions}\label{sec:junction}
The present section is devoted to establish the relationship existing between the
limit fields $\hat{\mathbf{u}}$ and $\check{\mathbf{u}}$. This is carried out by studying the junction conditions on $\Omega_{J}$.

\subsection{The Junction Conditions for Displacements}
From \eqref{eq:scalingpiastra} and \eqref{eq:scalingtrave}, the following equality must be satisfied for the identity of $\hat{\mathbf{u}}$ and $\check{\mathbf{u}}$ in $\Omega_{J\e}$:
\begin{equation*}
	\begin{aligned}
\hat{\mathbf{Q}}^{-1}_\e \hat{\mathbf{u}}_\e \circ \hat{\mathbf{q}}_\e^{-1} &= \check{\mathbf{Q}}^{-1}_\e \check{\mathbf{u}}_\e \circ \check{\mathbf{q}}_\e^{-1} & \textrm{ in } \Omega_{J\e},
\end{aligned}
\end{equation*}
which may be rewritten more explicitly for all  $(x_1, x_2, x_3) \in \Omega_{J}$ as
\begin{equation}\label{eq:cond}
\begin{split} \e^k \hat{u}_{1\e}(x_1,\e^w x_2, x_3)&= \e^k \check{u}_{1\e}(x_1,x_2, \e^{1-h}x_3),\\ 
\e^{k+w}\hat{u}_{2\e}(x_1,\e^w x_2,x_3)&= \e^k \check{u}_{2\e}(x_1,x_2, \e^{1-h} x_3), \\  
\e^{k} \hat{u}_{3\e}(x_1,\e^w x_2,x_3)&= \e^{k+1-h}\check{u}_{3\e}(x_1,x_2, \e^{1-h} x_3). \end{split}
\end{equation}
It is noteworthy that, in this way, the junction region, which originally
depends on $\e$, has been transformed into the fixed domain $\Omega_{J}$.

The following two technical Lemmata express an approximation of the trace operator; similar results can be found in \cite{ledret, Freddi2007}. We recall that $h<1$.
\begin{myLemma}{}\label{lemma:5.1}
	Let $w\in W^{1,2}(\Omega_{J})$ and $w_\e\in W^{1,2}(\Omega_{J})$ be a sequence such that $w_\e\weak w$ in $W^{1,2}(\Omega_{J})$. Then, the sequence of functions
	\begin{equation*}
		(x_1, x_2) \mapsto \Rint_{0}^T w_\e(x_1, x_2, \e^{1-h}x_3) \diff x_3
	\end{equation*}
converges in the norm of $L^2((-L,L)\times(-W,W))$ to the trace of the function $w$ on $(-L,L)\times(-W,W)\times\{0\}$.
The trace will be denoted simply by $w(x_1, x_2, 0)$.
\end{myLemma}
\begin{proof}
	We have:
	\begin{equation*}
		\begin{aligned}
			&\int_{-L}^{L}\int_{-W}^{W} \left| \Rint_{0}^T w_\e(x_1, x_2, \e^{1-h}x_3) - w_\e(x_1, x_2,0) \diff x_3  \right|^2 \diff x_2 \diff x_1\\
			&=\int_{-L}^{L}\int_{-W}^{W} \left| \Rint_{0}^{\e^{1-h}T} w_\e(x_1, x_2, t) - w_\e(x_1, x_2,0) \diff t  \right|^2 \diff x_2 \diff x_1\\
			&=\int_{-L}^{L}\int_{-W}^{W} \left| \Rint_{0}^{\e^{1-h}T} \int_0^t \partial_3 w_\e(x_1, x_2, s) \diff s \diff t  \right|^2 \diff x_2 \diff x_1\\
			&\leq \int_{-L}^{L}\int_{-W}^{W}  \Rint_{0}^{\e^{1-h}T} \left|\int_0^t \partial_3 w_\e(x_1, x_2, s) \diff s \right|^2\diff t   \diff x_2 \diff x_1\\
			&\leq \int_{-L}^{L}\int_{-W}^{W}  \Rint_{0}^{\e^{1-h}T} t \int_0^t  \left| \partial_3 w_\e(x_1, x_2, s) \right|^2 \diff s \diff t   \diff x_2 \diff x_1\\	
			&\lesssim  \e^{1-h} \int_{-L}^{L}\int_{-W}^{W}  \int_0^T  \left| \partial_3 w_\e(x_1, x_2, s) \right|^2 \diff s \diff x_2 \diff x_1 \lesssim \e^{1-h},\\			
		\end{aligned}
	\end{equation*}
where we have used a change of variable, the Fundamental Theorem of Calculus, and the Cauchy-Schwartz inequality (twice). The thesis follows from the continuity of the trace operator for functions in $W^{1,2}$ (see, for instance, \cite[Theorem 5.36]{Adams2003}) and from the general assumption $h<1$.
\end{proof}
With a similar argument, one can prove the following the counterpart for the stiffener of the previous Lemma.
\begin{myLemma}{}\label{lemma:5.1bis}
	Let $w\in W^{1,2}(\Omega_{J})$ and $w_\e\in W^{1,2}(\Omega_{J})$ be a sequence such that $w_\e\weak w$ in $W^{1,2}(\Omega_{J})$. Then, the sequence of functions
	\begin{equation*}
		(x_1, x_3) \mapsto \Rint_{-W}^{W} w_\e(x_1, \e^{w}x_2, x_3) \diff x_2
	\end{equation*}
	converges in the norm $L^2((-L,L)\times(0,T))$ to the trace of the function $w$ on $(-L,L)\times\{0\}\times(0,T)$. 
	The trace will be denoted simply by $w(x_1, 0, x_3)$.
 \end{myLemma}

The limit junction conditions for the  displacements are derived in the next result.
\begin{myLemma}\label{lemma:limitjoind}
	The following equalities hold for almost every $x_1\in(-L,L)$:
	\begin{equation}\label{eq:jcond}
		\begin{aligned}
			\hat{\varrho}_1 \left[\hat{\xi}_1(x_1,0) - \frac{T}{2}\partial_1\hat{\xi}_{3}(x_1,0)\right] &= \check{\varrho}_1 \check{\xi}_1(x_1),\\
			\hat{\varrho}_2 \left[\hat{\xi}_2(x_1,0) - \frac{T}{2}\partial_2\hat{\xi}_{3}(x_1,0)\right] &= \check{\varrho}_2 \check{\xi}_2(x_1),\\
			\hat{\varrho}_3 \hat{\xi}_3(x_1,0) &= \check{\varrho}_3 \check{\xi}_3(x_1),
		\end{aligned}
	\end{equation}
where $\hat{\varrho}_i\,:\,(0,1)\times\mathbb{R}^+\to\{0,1\}$ and 
$\check{\varrho}_i\,:\,(0,1)\times\mathbb{R}^+\to\{0,1\}$ are defined as follows:
\begin{equation*}
	\begin{aligned}
			\hat{\varrho}_1 &= \begin{cases}0, &\text{ if } k > 0,\\1, &\textrm{ if } k \leq 0,\end{cases} 			&\check{\varrho}_1 &= \begin{cases}0, &\textrm{ if } k < 0,\\1, &\textrm{ if } k \geq 0,\end{cases}\\
			\hat{\varrho}_2 &= \begin{cases}0, &\textrm{ if } k+w > 0,\\1, &\textrm{ if } k+w \leq 0,\end{cases} 			&\check{\varrho}_2 &= \begin{cases}0, &\textrm{ if } k+w < 0,\\1, &\textrm{ if } k+w \geq 0,\end{cases}\\
			\hat{\varrho}_3 &= \begin{cases}0, &\textrm{ if } k+h-1 > 0,\\1, &\textrm{ if } k+h-1 \leq 0,  \end{cases}
			&\check{\varrho}_3 &= \begin{cases}0, &\textrm{ if } k+h-1 < 0,\\1, &\textrm{ if } k+h-1 \geq 0.  \end{cases}\\
	\end{aligned}
\end{equation*}
\end{myLemma}
\begin{proof}
	We provide a detailed proof for the first component of the displacement, the others being similar.
	We consider the first of \eqref{eq:cond} averaged with respect to $x_2$ and $x_3$:
	\begin{equation}\label{medeq1}
		\e^k \Rint_{0}^{T} \Rint_{-W}^{W} \hat{u}_{1\e}(x_1,\e^w x_2, x_3) \diff x_2 \diff x_3 = 	 \Rint_{-W}^{W}\Rint_{0}^{T} \e^k \check{u}_{1\e}(x_1,x_2, \e^{1-h}x_3) \diff x_3 \diff x_2.
	\end{equation}
Since $\e^k	\check{u}_{1\e}\weak  \check{u}_{1}$ in $W^{1,2}(\Omega_J)$, by Lemma  \ref{lemma:5.1} it follows that
$$
 \Rint_{0}^{T} \e^k \check{u}_{1\e}(\cdot,\cdot, \e^{1-h}x_3)  \diff x_3  
 \to \check{u}_{1}(\cdot,\cdot, 0) \quad \mbox{in } L^2((-L,L)\times(-W,W)),
 $$
from which we deduce that
\begin{equation}\label{medeq2}
\Rint_{-W}^{W} \Rint_{0}^{T} \e^k \check{u}_{1\e}(\cdot,x_2, \e^{1-h}x_3)  \diff x_3  \diff x_2
 \to \Rint_{-W}^{W}\check{u}_{1}(\cdot,x_2,0)\diff x_2 \quad \mbox{in } L^2((-L,L)).
 \end{equation}
Similarly, since 	$\hat{u}_{1\e}\weak  \hat{u}_{1}$ in $W^{1,2}(\Omega_J)$, by Lemma  \ref{lemma:5.1bis} we find that
\begin{equation}\label{medeq3}
\Rint_{0}^{T} \Rint_{-W}^{W} \hat{u}_{1\e}(\cdot,\e^w x_2, x_3) \diff x_2 \diff x_3
 \to   \Rint_{0}^{T}\hat{u}_{1}(\cdot,0,x_3)\diff x_3 \quad \mbox{in } L^2((-L,L)).
\end{equation}

By passing to the limit in \eqref{medeq1}, taking into account \eqref{medeq2} and \eqref{medeq3}, we obtain 
\begin{equation}\label{medeq4}
\begin{aligned}
0&= \Rint_{-W}^{W}\check{u}_{1}(\cdot,x_2,0)\diff x_2  & &\quad \mbox{for }k>0,\\
\Rint_{0}^{T}\hat{u}_{1}(\cdot,0,x_3)\diff x_3 &= \Rint_{-W}^{W}\check{u}_{1}(\cdot,x_2,0)\diff x_2  & &\quad \mbox{for }k=0,\\
\Rint_{0}^{T}\hat{u}_{1}(\cdot,0,x_3)\diff x_3 &= 0  & &\quad \mbox{for }k<0.
\end{aligned}
\end{equation}
Since
\begin{equation*}
\begin{aligned}
\Rint_{-W}^{W} \check{u}_{1}(x_1, x_2, 0) \diff x_2 &= \Rint_{-W}^{W} \left(\check{\xi}_1(x_1) - x_2\partial_1\check{\xi}_{2}(x_1)\right) \diff x_2 = \check{\xi}_1(x_1),\\
\Rint_{0}^{T} \hat{u}_{1}(x_1, 0, x_3) \diff x_3 &= \Rint_{0}^{T} \left(\hat{\xi}_1(x_1,0) - x_3\partial_1\hat{\xi}_{3}(x_1,0)\right) \diff x_3 = \hat{\xi}_1(x_1,0) - \frac{T}{2}\partial_1\hat{\xi}_{3}(x_1,0),
\end{aligned}
\end{equation*}
equation \eqref{medeq4} implies the first identity of \eqref{eq:jcond}.

	For the other components of the displacement, from \eqref{eq:cond}, we find:
\begin{equation*}
	\e^{k+w}\Rint_{0}^{T} \Rint_{-W}^{W}  \hat{u}_{2\e}(x_1,\e^w x_2, x_3) \diff x_2 \diff x_3 = 	 \Rint_{-W}^{W}\Rint_{0}^{T} \e^k \check{u}_{2\e}(x_1,x_2, \e^{1-h}x_3) \diff x_2 \diff x_3,
\end{equation*}

\begin{equation*}
	 \e^{k+h-1}\Rint_{0}^{T} \Rint_{-W}^{W} \hat{u}_{3\e}(x_1,\e^w x_2, x_3) \diff x_2 \diff x_3 = 	 \Rint_{-W}^{W}\Rint_{0}^{T} \e^k \check{u}_{3\e}(x_1,x_2, \e^{1-h}x_3) \diff x_2 \diff x_3,
\end{equation*}
and arguing as above we complete the proof.
\end{proof}

\subsection{The Junction Conditions for the Torsion Angle}\label{sec:junctionangle}

This Section is dedicated to establish a further relationship existing between the limit fields, involving the torsion angle of the stiffener.  
Hereafter, $M\coloneqq\max\{w,h\}$, $m\coloneqq\min\{w,h\}$.
As a trivial consequence, $M\geq m$, with equality holding if and only if $h=w$.

We start with a scaled Korn-type inequality.
\begin{myLemma}\label{lemma:abp}
Let $\mathbf{u}\in W^{1,2}(\check{\Omega}_\e, \mathbb{R}^{3})$. Then, we have
\begin{equation*}
\norm{\mathbf{u}}_{W^{1,2}(\check{\Omega}_\e,\mathbb{R}^3)}\lesssim \frac{1}{\e^{M}}\norm{\mathbf{E}\mathbf{u}}_{L^2(\check{\Omega}_\e,\mathbb{R}^{3 \times 3})}.
\end{equation*}
\end{myLemma}
\begin{proof}
We use an argument from \cite[Theorem 3.2]{Freddi2013}.
Decompose $\check{\Omega}_\e$ in parallelepipeds having cross-section $\e^{M}\times\e^{M}$. For each of them, apply the scaled Korn inequality of the type in \cite[Theorem A.1.]{anzellotti} (see also \cite[Theorem 2]{Kondratev1989}). Then, summing up, the thesis follows.
%\QEDB
\end{proof}

\begin{myLemma}\label{lemmatheta}
With the assumptions and the notation of Lemma \ref{compattezzatrave}, we have
\begin{equation}\label{HE}
\norm{\e^{k+M}\check{\mathbf{H}}_\e \check{\mathbf{u}}_\e}_{L^2(\check{\Omega},\mathbb{R}^{3 \times 3})}\lesssim\norm{\e^k \check{\mathbf{E}}_\e\check{\mathbf{u}}_\e}_{L^2(\check{\Omega},\mathbb{R}^{3 \times 3})} < \infty.
\end{equation}
In particular, up to a subsequence,
$$
\e^{k+M}\check{\mathbf{H}}_\e \check{\mathbf{u}}_\e \weak \check{\mathbf{H}}
\qquad\mbox{in }L^2(\check{\Omega}, \mathbb{R}^{3 \times 3}),
$$
where $\check{\mathbf{H}}\in L^2(\check{\Omega}, \Skw(\mathbb{R}^{3 \times 3}))$ has components
$$
\check H_{12}=\begin{cases} \partial_2\check{u}_1, &\textrm{ if } M=w,\\
0, &\textrm{ otherwise}, \end{cases}
\qquad
\check H_{13}=\begin{cases} \partial_3\check{u}_1, &\textrm{ if } M=h,\\
0, &\textrm{ otherwise }, \end{cases}
\qquad
\check H_{32}=\vartheta,$$
with $\vartheta$  a function in $L^2(\check{\Omega})$.
\end{myLemma}

\begin{proof}
The proof of \eqref{HE} follows immediately by changing variables and scaling into the fixed domain the result of Lemma \ref{lemma:abp},  and by recalling Lemma \ref{compattezzatrave}.
From this bound we deduce that there exists a subsequence, not relabeled, of 
$\e^{k+M}\check{\mathbf{H}}_\e \check{\mathbf{u}}_\e$ and a
$\check{\mathbf{H}}\in L^2(\check{\Omega}, \mathbb{R}^{3 \times 3})$, such that 
$$
\e^{k+M}\check{\mathbf{H}}_\e \check{\mathbf{u}}_\e \weak \check{\mathbf{H}}\quad \mbox{in } L^2(\check{\Omega}, \mathbb{R}^{3 \times 3}).
$$
Since $\e^{M}\left(\e^k \check{\mathbf{E}}_\e \check{\mathbf{u}}_\e\right)\to 0$ in $L^2(\check{\Omega}, \mathbb{R}^{3 \times 3})$, it follows that $\e^{k+M} \check{\mathbf{W}}_\e \check{\mathbf{u}}_\e \weak \check{\mathbf{H}}$ in $L^2(\check{\Omega}, \mathbb{R}^{3 \times 3})$. Accordingly, $\check{\mathbf{H}} $ is a skew-symmetric tensor field. Moreover, from the definition \eqref{eq:Hcheck} and the convergence of
$\e^k\check{\mathbf{u}}_{\e} \weak \check{\mathbf{u}}$ in $W^{1,2}(\check{\Omega},\mathbb{R}^3)$, we deduce that
\begin{equation}\label{eq:seqacaso20}
\begin{split}
\e^{k+M}\left( \check{\mathbf{H}}_\e \check{\mathbf{u}}_\e\right)_{12} &= \e^{k+M-w} \partial_2 \check{u}_{1\e}\weak \begin{cases} \partial_2\check{u}_1, &\textrm{ if } M=w,\\
0, &\textrm{ otherwise}, \end{cases}\\
\e^{k+M}\left( \check{\mathbf{H}}_\e \check{\mathbf{u}}_\e\right)_{13} &= \e^{k+M-h} \partial_3 \check{u}_{1\e}\weak \begin{cases} \partial_3\check{u}_1, &\textrm{ if } M=h,\\
0, &\textrm{ otherwise}, 
\end{cases}
\end{split}
\end{equation}
The proof is completed by setting $\vartheta\coloneqq \check H_{32}$.
\end{proof}

The function $\vartheta$ defined in Lemma \ref{lemmatheta} can be interpreted as the rotation angle, or torsion angle, of the stiffener cross-section around the longitudinal axis $x_1$. 
It is noteworthy that if $w=h$ the first two sequences of (\ref{eq:seqacaso20}) have non-trivial limits simultaneously.

We shall now characterize the torsion angle $\vartheta$. The idea is to show that the cross-sectional displacement field of the stiffener can be approximated by a rigid one.
To do so, let us introduce the set of infinitesimal rigid displacements on $\check{\omega}$:
\begin{equation*}
\mathfrak{R}(\check{\omega})\coloneqq\{ \mathbf{r} \in L^2(\check{\omega}, \mathbb{R}^2):\exists \varphi\in \mathbb{R},\ \mathbf{c}\in \mathbb{R}^2,\ r_a(x_2, x_3)=\mathcal{E}_{ba}x_b \varphi + c_a\},
\end{equation*}
where $\mathcal{E}$ is the Ricci symbol ($\mathcal{E}_{22}=\mathcal{E}_{33}=0$, $\mathcal{E}_{23}= -\mathcal{E}_{32}=1$). Then, $\mathfrak{R}(\check{\omega})$ is a finite closed subspace of $W^{1,2}(\check{\omega}, \mathbb{R}^2)$.
 We indicate with $\mathcal{P}$ the projection of $L^2(\check{\omega}, \mathbb{R}^2)$ onto $\mathfrak{R}(\check{\omega})$.
It can be shown (see \cite[Theorem 2.5]{oleinik}) that a two-dimensional Korn inequality holds for all functions $\mathbf{w}\in W^{1,2}(\check{\omega}, \mathbb{R}^2)$:
\begin{equation}\label{kornproj}
\norm{\mathbf{w}-\mathcal{P}\mathbf{w} }_{W^{1,2}(\check{\omega}, \mathbb{R}^2)}\lesssim \norm{\mathbf{Ew}}_{L^2(\check{\omega}, \mathbb{R}^{2\times 2})}.
\end{equation}
Denoting with $(x_2(G), x_3(G))=(0, \frac{H}{2})$ the coordinates of the centre of mass of $\check{\omega}$, we define (with summation convention and $a,b,c,d\in\{2,3\}$):
\begin{equation}\label{eq:disprigidi}
\begin{aligned}
t_a(\check{\mathbf{u}}_\e) &\coloneqq\Rint_{\check{\omega}} \check{u}_{a\e} \diff x_2\diff x_3,
&I_G(\check{\omega}) &\coloneqq\int_{\check{\omega}} (x_2 - x_2(G))^2+(x_3 - x_3(G))^2 \diff x_2\diff x_3,\\
\vartheta_\e (x_1)&\coloneqq\frac{1}{I_G}\int_{\check{\omega}} \mathcal{E}_{cd}\,(x_c - x_c(G))\,\check{u}_{d\e}\, \diff x_2\,\diff x_3,
&\left(\mathcal{P}\check{\mathbf{u}}_\e\right)_a &\coloneqq t_a(\check{\mathbf{u}}_\e)+\mathcal{E}_{ba}(x_b-x_b(G))\vartheta_\e(x_1).
\end{aligned}
\end{equation}

\begin{myLemma}\label{lemma53}
Let $\e^k \check{\mathbf{u}}_\e$ be a sequence satisfying the assumptions of Lemma \ref{lemmatheta}. Then,
\begin{equation*}
\norm{ \e^{k} \left(\check{\mathbf{u}}_\e-\mathcal{P}\check{\mathbf{u}}_\e\right)}_{L^2((-L,L), W^{1,2}(\check{\omega}, \mathbb{R}^{2}))} \lesssim \e^{2m}.
\end{equation*}
\end{myLemma}
\begin{proof}

Taking into account (\ref{kornproj}) and \eqref{eq:straintrave}, we have
\begin{equation*}
\begin{split}
\int_{-L}^{L} \norm{  \e^{k} \left(\check{\mathbf{u}}_\e-\mathcal{P}\check{\mathbf{u}}_\e\right)}^2_{W^{1,2}(\check{\omega}, \mathbb{R}^{2})}\diff x_1  & \lesssim\int_{-L}^{L} \sum_{ab} \norm{ \e^{k}\left(\mathbf{E} \check{\mathbf{u}}_\e \right)_{ab}}^2_{L^2(\check{\omega})}\diff x_1\\
&\lesssim \sum_{ab} \norm{  \e^{k} (\check{\mathbf{Q}}_\e \check{\mathbf{E}}_\e \check{\mathbf{u}}_\e \check{\mathbf{Q}}_\e)_{ab}}^2_{L^2(\check{\Omega})}\\
&\begin{split} \lesssim \norm{  \e^{k+2w} (\check{\mathbf{E}}_\e \check{\mathbf{u}}_\e)_{22}}^2_{L^2(\check{\Omega})}
&+  \norm{  \e^{k+2h} (\check{\mathbf{E}}_\e \check{\mathbf{u}}_\e)_{33}}^2_{L^2(\check{\Omega})}\\
&+ \norm{  \e^{k+w+h} (\check{\mathbf{E}}_\e \check{\mathbf{u}}_\e)_{23}}^2_{L^2(\check{\Omega})} \end{split}\\
& \lesssim  \e^{4m} +  \e^{4M} +   \e^{2(M+m)}\\
& \lesssim \e^{4m},
\end{split}
\end{equation*}
where we have used the fact that one among $w$ and $h$ equals $M$ (while the other, by definition, equals $m$), and the fact that $\e^{2M} \leq \e^{M+m} \leq \e^{2m}$.
\end{proof} 

\begin{myLemma}\label{convteta}
With the assumptions and notation of Lemmata \ref{lemmatheta} and \ref{lemma53}, we have
\begin{enumerate}
\item $\e^{k-m}\vartheta_\e \weak \vartheta$ in $L^2(\check{\Omega})$;\label{c1} 
\item $\vartheta$ is a function of $x_1$ only;\label{c2}
\item $\vartheta\in W^{1,2}_0((-L,L))$,\label{c3}
\end{enumerate}
where $\vartheta$ and $\vartheta_\e$ are defined in \eqref{eq:seqacaso20} and  \eqref{eq:disprigidi}, respectively.
\end{myLemma}

\begin{proof}
From Lemma~\ref{lemma53}, we have
\begin{equation}\label{eqw}
\norm{ \e^{k} \partial_a(\check{\mathbf{u}}_\e-\mathcal{P}\check{\mathbf{u}}_\e)}_{L^2(\check{\Omega})} \lesssim \e^{2m}.
\end{equation}
Since $(\mathbf{W}\mathcal{P}\check{\mathbf{u}}_\e)_{23}=-\vartheta_\e$ by \eqref{eq:disprigidi},  it follows that $
\norm{ \e^{k} \vartheta_\e + \e^{k}\left(\mathbf{W}\check{\mathbf{u}}_\e\right)_{23}}_{L^2(\check{\Omega})}  = 
\norm{ 
\e^{k} \left(\mathbf{W} \left(\check{\mathbf{u}}_\e-\mathcal{P}\check{\mathbf{u}}_\e\right)\right)_{23}
}_{L^2(\check{\Omega})}\lesssim\e^{2m},
$
thanks to Lemma \ref{lemma53}.

By using the fact that $M+m=h+w$, we can rewrite this inequality as
$
	\norm{ \e^{k} \vartheta_\e + \e^{k+M+m} \left(\check{\mathbf{W}}_\e \check{\mathbf{u}}_\e\right)_{23}}_{L^2(\check{\Omega})}	\lesssim \e^{2m},
$
from which we deduce
$
	\norm{ \e^{k-m} \vartheta_\e + \e^{k+M} \left(\check{\mathbf{W}}_\e \check{\mathbf{u}}_\e\right)_{23}}_{L^2(\check{\Omega})}	\lesssim \e^{m}.
$
Recalling \eqref{eq:seqacaso20}, we conclude that $\e^{k-m} \vartheta_\e \weak \vartheta$ in $L^2(\check{\Omega})$, thus proving claim \ref{c1}.\\
Claim \ref{c2} follows from claim \ref{c1}, since $\vartheta_\e$, by definition, does not depend on $x_2$ and $x_3$. \\
To prove claim \ref{c3} we consider a test function $\psi\in C_0^\infty(\check{\omega})$, such that
$
\int_{\check{\omega}}\psi \diff x_2\diff x_3 =-\frac{I_G(\check{\omega})}{2}.
$
Taking into account (\ref{eq:disprigidi}), we find (with summation convention):
\begin{equation*}
\begin{split}
 I_G(\check{\omega})\e^{k-m}\vartheta_\e&=-2\e^{k-m}\vartheta_\e\int_{\check{\omega}}\psi \diff x_2\diff x_3= -\vartheta_\e\e^{k-m}\int_{\check{\omega}}\psi \partial_a \left(x_a - x_a(G)\right) \diff x_2\diff x_3\\
&=\vartheta_\e\e^{k-m}\int_{\check{\omega}}\partial_a \psi  \left(x_a - x_a(G)\right) \diff x_2\diff x_3=\e^{k-m}\int_{\check{\omega}}\mathcal{E}_{ac} \partial_a \psi \left(\mathcal{E}_{bc} \left(x_b - x_b(G)\right) \vartheta_\e\right) \diff x_2\diff x_3\\
&=\e^{k-m}\int_{\check{\omega}}\mathcal{E}_{ac} \partial_a\psi \left((\mathcal{P}\check{\mathbf{u}}_\e)_c-t_c\right) \diff x_2\diff x_3\\
&=\e^{k-m}\int_{\check{\omega}}\mathcal{E}_{ac} \partial_a\psi (\mathcal{P}\check{\mathbf{u}}_\e)_c \diff x_2\diff x_3 - \e^{k-m}\int_{\check{\omega}}\mathcal{E}_{ac}\partial_a\psi \left(\Rint_{\check{\omega}} \check{u}_{c\e}\diff x_2\diff x_3\right)  \diff x_2\diff x_3\\
&=\e^{k-m}\int_{\check{\omega}}\mathcal{E}_{ac} \partial_a\psi (\mathcal{P}\check{\mathbf{u}}_\e)_c \diff x_2\diff x_3\\
&=\e^{k-m}\int_{\check{\omega}}\mathcal{E}_{ac} \partial_a\psi \check{u}_{c\e} \diff x_2\diff x_3-\e^{k-m}\int_{\check{\omega}}\mathcal{E}_{ac} \partial_a\psi \left(\check{\mathbf{u}}_\e-\mathcal{P}\check{\mathbf{u}}_\e\right)_c \diff x_2\diff x_3.\\
\end{split}
\end{equation*}
Setting
$
\tilde{\vartheta}_\e\coloneqq\frac{\e^{k-m}}{I_G}\int_{\check{\omega}}\mathcal{E}_{ac} \partial_a\psi \check{u}_{c\e} \diff x_2\diff x_3
$
and taking into account inequality (\ref{eqw}), we deduce 
\begin{equation}\label{eqe}
\e^{k-m} \vartheta_\e - \tilde{\vartheta}_\e\rightarrow 0\hspace{1cm} \textrm{ in } L^2(\check{\Omega}).
\end{equation}
We now show that $\partial_1\tilde{\vartheta}_\e$ is bounded in $L^2$. By definition, $\mathcal{E}_{ab}\partial_a \partial_b \psi =0$ everywhere in $\check{\omega}$ and $\partial_a \psi =0$ on $\partial\check{\omega}$; therefore, we have
\begin{equation*}
\begin{split}
I_G \partial_1\tilde{\vartheta}_\e &=\e^{k-m}\int_{\check{\omega}}\mathcal{E}_{ac} \partial_a\psi \partial_1 \check{u}_{c\e} \diff x_2\diff x_3 \\
&=2\e^{k-m}\int_{\check{\omega}}\mathcal{E}_{ac} \partial_a\psi (\mathbf{E} \check{\mathbf{u}}_\e)_{c 1}\diff x_2\diff x_3 -\e^{k-m}\int_{\check{\omega}}\mathcal{E}_{ac} \partial_a\psi \partial_c \check{u}_{1\e}\diff x_2\diff x_3\\
&=2\e^{k-m}\int_{\check{\omega}}\mathcal{E}_{ac} \partial_a\psi (\mathbf{E} \check{\mathbf{u}}_\e)_{c 1}\diff x_2\diff x_3 -\e^{k-m}\int_{\check{\omega}}\partial_c\left(\mathcal{E}_{ac} \partial_a\psi  \check{u}_{1\e}\right)\diff x_2\diff x_3
+\e^{k-m}\int_{\check{\omega}}\mathcal{E}_{ac} \partial_a\partial_c\psi  \check{u}_{1\e}\diff x_2\diff x_3
\\
&=2\e^{k-m}\int_{\check{\omega}}\mathcal{E}_{ac} \partial_a\psi (\mathbf{E} \check{\mathbf{u}}_\e)_{c 1}\diff x_2\diff x_3 -\e^{k-m}\int_{\partial\check{\omega}} \mathcal{E}_{ac}\partial_a\psi n_c \check{u}_{1\e} \diff s
+\e^{k-m}\int_{\check{\omega}}\mathcal{E}_{ac} \partial_a\partial_c\psi  \check{u}_{1\e}\diff x_2\diff x_3
\\
&=2\e^{k-m}\int_{\check{\omega}}\mathcal{E}_{ac} \partial_a\psi (\mathbf{E} \check{\mathbf{u}}_\e)_{c 1}\diff x_2\diff x_3\\
&=2\e^{h-m}\int_{\check{\omega}}\partial_2\psi \e^k(\check{\mathbf{E}}_\e \check{\mathbf{u}}_\e)_{31}\diff x_2\diff x_3 - 2\e^{w-m}\int_{\check{\omega}}\partial_3\psi \e^k(\check{\mathbf{E}}_\e \check{\mathbf{u}}_\e)_{21}\diff x_2\diff x_3,
\end{split}
\end{equation*}
since, by \eqref{eq:straintrave}, we have $(\mathbf{E} \check{\mathbf{u}}_\e)_{2 1}=\e^{w}(\check{\mathbf{E}}_\e \check{\mathbf{u}}_\e)_{21} \; \text{and} \; (\mathbf{E} \check{\mathbf{u}}_\e)_{3 1}=\e^{h}(\check{\mathbf{E}}_\e \check{\mathbf{u}}_\e)_{31}$.
Hence, by noticing that one among $w$ or $h$ equals $m$, the remaining one being equal to $M$, recalling Lemma \ref{compattezzatrave} we
deduce that $\partial_1\tilde{\vartheta}_\e$ is bounded in $L^2((-L,L))$. From (\ref{eqe}) and Lemma~\ref{convteta} we conclude that 
$\tilde{\vartheta}_\e \weak \vartheta$ in  $W^{1,2}(\check{\Omega})$.
Since, additionally,  $\tilde{\vartheta}_\e(L)=0$, we conclude that $\tilde{\vartheta}\in W^{1,2}_0((-L,L))$, and hence claim \ref{c3} is proved.
%\QEDB
\end{proof}

Finally, we derive the last junction condition involving the torsion angle $\vartheta$. 
\begin{myLemma}\label{lemma:limitjoint}
	With the assumptions and notation of Lemma \ref{convteta}, the following equality holds for almost every $x_1\in(-L,L)$
	\begin{equation}\label{eq:lastjc}
		\hat{\varrho}_\vartheta\partial_2\hat{\xi}_3(x_1,0) = \check{\varrho}_\vartheta\vartheta(x_1),
	\end{equation}
where the functions $\hat{\varrho}_\vartheta\,:\,(0,1)\times\mathbb{R}^+\to\{0,1\}$ and $\check{\varrho}_\vartheta\,:\,(0,1)\times\mathbb{R}^+\to\{0,1\}$ are defined by
\begin{equation*}
	\begin{aligned}
		\hat{\varrho}_\vartheta &\coloneqq 
		\begin{cases}
			0, &\textrm{ if } k+M-1 > 0,\\
			1, &\textrm{ if } k+M-1 \leq 0,
		\end{cases}
		&\check{\varrho}_\vartheta &\coloneqq 
		\begin{cases}
			0, &\textrm{ if } k+M-1 < 0,\\
			1, &\textrm{ if } k+M-1 \geq 0.
		\end{cases}
	\end{aligned}
\end{equation*}
\end{myLemma}
\begin{proof}
Taking the derivative with respect to $x_3$ of the second of (\ref{eq:cond}), one finds
\begin{equation}\label{eq:boh}
	\begin{aligned}
		\e^{k+w}\partial_3\hat{u}_{2\e}(x_1, \e^w x_2, x_3) &= \e^{k+1-h} \partial_3\check{u}_{2\e}(x_1,x_2,\e^{1-h}x_3) & \forall \ (x_1, x_2, x_3)\in \Omega_{J}.
	\end{aligned}
\end{equation}
By multiplying each member by $\e^{-m}$, one obtains
\begin{equation*}
	\begin{aligned}
		\e^{k+w-m}\partial_3\hat{u}_{2\e}(x_1, \e^w x_2, x_3) &= \e^{k+1-h-m} \partial_3\check{u}_{2\e}(x_1,x_2,\e^{1-h}x_3) & \forall \ (x_1, x_2, x_3)\in \Omega_{J},
	\end{aligned}
\end{equation*}
which can be rearranged as
\begin{equation}\label{eq:boh2}
	\begin{aligned}
		\e^{k+w-m}\partial_3\hat{u}_{2\e}(x_1, \e^w x_2, x_3) = \e^{k+1-h+M} 2(\check{\mathbf{E}}_\e \check{\mathbf{u}}_\e)_{32}(x_1,x_2,\e^{1-h}x_3) - \e^{k+1-h-m} \partial_2 \check{u}_{3\e}(x_1,x_2,\e^{1-h}x_3)
	\end{aligned}
\end{equation}
for every $ \ (x_1, x_2, x_3)\in \Omega_{J}$. In deriving \eqref{eq:boh2}, we have used the identity $(\check{\mathbf{E}} \check{\mathbf{u}}_\e)_{32} = \e^{h+w}(\check{\mathbf{E}}_\e \check{\mathbf{u}}_\e)_{32}$ (see \eqref{eq:straintrave}) and the identity $k+1-h-m+h+w =k+1-h+M$ since, by definition, $h+w=M+m$.\\
We now use some arguments that can be found in the monograph by Le Dret \cite[Section 4]{ledret}.\\
From Lemma \ref{compattezzapiastra}, we know that $\partial_3 \hat{u}_{2\e}$ is bounded in $L^2(\hat{\Omega})$, which implies that $\partial_3 \hat{u}_{2\e}$ is bounded in $L^2((-W,W),  L^2(\omega_{13}))$, where we have set $\omega_{13}\coloneqq (-L,L)\times(0,T)$. Similarly, from Lemma \ref{lemmatheta} and \eqref{eq:seqacaso20}, we know that $\e^{k-m}\partial_2 \check{u}_{3\e}$ is bounded in $ L^2(\check{\Omega})$, which implies that $\e^{k-m}\partial_2 \check{u}_{3\e}$ is bounded in $ L^2((0,T),  L^2(\omega_{12}))$, where we have set $\omega_{12}\coloneqq (-L,L)\times(-W,W)$.\\
We now prove that $\partial_2\partial_3 \hat{u}_{2\e}$ is bounded in $ L^2((-W,W), W^{-1,2}(\omega_{13}))$ and that $\e^{k-m}\partial_3\partial_2 \check{u}_{3\e}$ is bounded in $ L^2((0,T), W^{-1,2}(\omega_{12}))$.\\
For every function $\psi_0\in W^{1,2}_0(\omega_{13})$, we have
\begin{equation*}
	\begin{aligned}
	\left|<\partial_2\partial_3 \hat{u}_{2\e},\psi_0> \right| &= %\left|\int_{\omega_{13}} \partial_2\partial_3 \hat{u}_{2\e} \psi_0 \diff x_1 \diff x_3 \right| = 
	\left|-\int_{\omega_{13}} \partial_2 \hat{u}_{2\e} \partial_3\psi_0 \diff x_1 \diff x_3 \right|
\leq \norm{\partial_2 \hat{u}_{2\e} }_{L^2(\omega_{13})}  \norm{\partial_3 \psi_0}_{L^2(\omega_{13})}\\
	&= \norm{\left(\hat{\mathbf{E}}_\e \hat{\mathbf{u}}_\e \right)_{22} }_{L^2(\omega_{13})}  \norm{\partial_3 \psi_0}_{L^2(\omega_{13})}
	\lesssim  \norm{\left(\hat{\mathbf{E}}_\e \hat{\mathbf{u}}_\e \right)_{22} }_{L^2(\omega_{13})} \norm{\psi_0}_{W^{1,2}_0(\omega_{13})}
	\end{aligned}
\end{equation*}
hence
$$
\norm{\partial_2\partial_3 \hat{u}_{2\e} }_{W^{-1,2}(\omega_{13})}\lesssim  \norm{\left(\hat{\mathbf{E}}_\e \hat{\mathbf{u}}_\e \right)_{22} }_{L^2(\omega_{13})}
$$
and
$$
\sup_\e \int_{-W}^W \norm{\partial_2\partial_3 \hat{u}_{2\e} }_{W^{-1,2}(\omega_{13})}^2
\diff x_2\lesssim \sup_\e \int_{-W}^W \norm{\left(\hat{\mathbf{E}}_\e \hat{\mathbf{u}}_\e \right)_{22} }_{L^2(\omega_{13})}^2
\diff x_2<\infty,
$$
where we have used the compactness result of Lemma \ref{compattezzapiastra}.
Hence, $\partial_2\partial_3 \hat{u}_{2\e}$ is bounded in $ L^2((-W,W), W^{-1,2}(\omega_{13}))$. \\
Similarly, for every function $\psi_0\in W^{1,2}_0(\omega_{12})$, we have
\begin{equation*}
	\begin{aligned}
		\left|<\e^{k-m}\partial_3\partial_2 \check{u}_{3\e},\psi_0> \right| &%= \left|\int_{\omega_{12}} \e^{k-m}\partial_3\partial_2 \check{u}_{3\e} \psi_0 \diff x_1 \diff x_2 \right| 
		= \left|-\int_{\omega_{12}} \e^{k-m} \partial_3 \check{u}_{3\e} \partial_2\psi_0 \diff x_1 \diff x_2 \right|\\
		&\leq \norm{\e^{k-m}\partial_3 \check{u}_{3\e} }_{L^2(\omega_{12})}  \norm{\partial_2 \psi_0}_{L^2(\omega_{12})}\\
		&= \e^{-m+2h} \norm{\e^{k}\left(\check{\mathbf{E}}_\e \check{\mathbf{u}}_\e \right)_{33} }_{L^2(\omega_{12})}  \norm{\partial_2 \psi_0}_{L^2(\omega_{12})},%\\
		%& \lesssim \e^{-m+2h}\norm{\psi_0}_{W^{1,2}_0(\omega_{12})} < \infty,
	\end{aligned}
\end{equation*}
where we have used \eqref{eq:straintrave}. Arguing as for $\partial_2\partial_3 \hat{u}_{2\e}$, and using the compactness result of Lemma \ref{compattezzatrave} we deduce that $\partial_3\partial_2 \check{u}_{3\e}$ is bounded in $ L^2((0,T), W^{-1,2}(\omega_{12}))$, since either $-m+2h = h$ ($>0$) or $-m+2h = 2M-m$ ($ > 0$). \\

To sum up, we have
\begin{equation}\label{eq:pluto}
	\begin{aligned}
	\partial_3 \hat{u}_{2\e} &\textrm{ is bounded in } W^{1,2}((-W,W), W^{-1,2}(\omega_{13})),
	&\e^{k-m} \partial_2 \check{u}_{3\e} &\textrm{ is bounded in } W^{1,2}((0,T), W^{-1,2}(\omega_{12})).\\
\end{aligned}
\end{equation}
From \cite[Lemma 1.3]{ledret}, we have that $W^{1,2}((-W,W), X)$, with $X$ any separable Hilbert space, is continuously embedded in the Holder space $C^{0,1/2}((-W,W), X)$. Thus, the following estimates (which represent a generalization of Lemmata \ref{lemma:5.1} and \ref{lemma:5.1bis})
hold
\begin{equation}\label{eq:ledretext}
	\begin{aligned}
	\norm{\partial_3 \hat{u}_{2\e} (x_1, \e^w x_2, x_3) - \partial_3 \hat{u}_{2\e} (x_1, 0, x_3)}^2_{W^{-1,2}(\omega_{13})} &\lesssim \e^w \norm{\partial_3 \hat{u}_{2\e}}^2_{W^{1,2}((-W,W), W^{-1,2}(\omega_{13}))}\\
		\norm{\e^{k-m}\partial_2 \check{u}_{3\e} (x_1, x_2, \e^{1-h} x_3) - \e^{k-m}\partial_2 \check{u}_{3\e} (x_1, x_2, 0)}^2_{W^{-1,2}(\omega_{12})} &\lesssim \e^{1-h} \norm{\e^{k-m}\partial_2 \check{u}_{3\e}}^2_{W^{1,2}((0,T),  W^{-1,2}(\omega_{12}))}.
\end{aligned}
\end{equation}
We multiply both members of \eqref{eq:boh2} by three functions $\varphi_i$, whose variable is $x_i$,  such that $\prod_{i=1}^{3} \varphi_i \in C_0^\infty(\Omega_{J})$, and we integrate over $\Omega_{J}$:
\begin{equation}\label{eq:boh30}
	\begin{aligned}
	& \mbox{LHS}_\e\coloneqq \e^{k+w-m}\int_{\Omega_{J}} \partial_3 \hat{u}_{2\e}(x_1,\e^w x_2, x_3)\prod_{i=1}^{3} \varphi_i(x_i) \diff x_1 \diff x_2 \diff x_3 \\
	& \hspace{3cm}= 2\e^{M+1-h} \int_{\Omega_{J}} \e^{k} (\check{\mathbf{E}}_\e \check{\mathbf{u}}_\e)_{32}(x_1,x_2,\e^{1-h}x_3)\prod_{i=1}^{3} \varphi_i(x_i) \diff x_1 \diff x_2 \diff x_3 \\
	& \hspace{5cm}- \int_{\Omega_{J}} \e^{k+1-h-m} \partial_2\check{u}_{3\e} (x_1,x_2,\e^{1-h}x_3) \prod_{i=1}^{3} \varphi_i(x_i) \diff x_1 \diff x_2 \diff x_3=:\mbox{RHS}_\e.
		\end{aligned}
\end{equation}

The left-hand side of \eqref{eq:boh30} can be rearranged as
\begin{equation*}%\label{eq:boh3}
	\begin{aligned}
	\mbox{LHS}_\e&= -\e^{k+w-m}\int_{-W}^{W} \varphi_2 \left(\int_{\omega_{13}}  \hat{u}_{2\e}(x_1,\e^w x_2, x_3)\varphi_1 \partial_3\varphi_3 \diff x_1  \diff x_3 \right) \diff x_2\\
		&= -\e^{k+w-m}\int_{-W}^{W} \varphi_2 \left(\int_{\omega_{13}}  \hat{u}_{2\e}(x_1,0, x_3)\varphi_1 \partial_3\varphi_3 \diff x_1  \diff x_3 \right) \diff x_2\\
		& \hspace{3cm} +\e^{k+w-m}\int_{-W}^{W} \varphi_2 < \partial_3 \hat{u}_{2\e}(x_1,\e^w x_2, x_3)-\partial_3 \hat{u}_{2\e}(x_1,0, x_3),\varphi_1 \varphi_3 > \diff x_2
	\end{aligned}
\end{equation*}
from which we deduce that
\begin{equation*}%\label{eq:boh300}
	\begin{aligned}
	 & \left|\mbox{LHS}_\e+\e^{k+w-m}\int_{-W}^{W} \varphi_2 \left(\int_{\omega_{13}}  \hat{u}_{2\e}(x_1,0, x_3)\varphi_1 \partial_3\varphi_3 \diff x_1  \diff x_3 \right) \diff x_2 \right|\\
		& \hspace{3cm} \le\e^{k+w-m}\int_{-W}^{W} \left|\varphi_2\right| \|\partial_3 \hat{u}_{2\e}(x_1,\e^w x_2, x_3)-\partial_3 \hat{u}_{2\e}(x_1,0, x_3)\|_{W^{-1,2}(\omega_{13})}\|\varphi_1 \varphi_3\|_{W^{1,2}_0} \diff x_2\\
		& \hspace{3cm} \lesssim \e^{k+w-m+w/2}
	\end{aligned}
\end{equation*}
where the last inequality follows by using \eqref{eq:ledretext} and \eqref{eq:pluto}.
Thus,
\begin{equation}\label{eq:boh300}
	\begin{aligned}
	\frac{\e^m  \mbox{LHS}_\e}{\e^{k+w}}=&-\int_{-W}^{W} \varphi_2 \left(\int_{\omega_{13}}  \hat{u}_{2\e}(x_1,0, x_3)\varphi_1 \partial_3\varphi_3 \diff x_1  \diff x_3 \right) \diff x_2 +O( \e^{w/2})\\
	& \to -\int_{-W}^{W} \varphi_2 \left(\int_{\omega_{13}}  \hat{u}_{2}(x_1,0, x_3)\varphi_1 \partial_3\varphi_3 \diff x_1  \diff x_3 \right) \diff x_2\\
	&\hspace{4cm}
	=\int_{\Omega_{J}} -\partial_2 \hat{\xi}_3(x_1,0)\prod_{i=1}^{3} \varphi_i(x_i) \diff x_1 \diff x_2 \diff x_3, \\
	\end{aligned}	
\end{equation}
since $\partial_3 \hat{u}_2(x_1,0, x_3) = -\partial_2 \hat{\xi}_3(x_1,0)$.

The first integral of the right-hand side of \eqref{eq:boh30} is of order $O(\e^{M+1-h})$,  by the compactness result of Lemma \ref{compattezzatrave}, while the second integral
can be handled as the left-hand side of \eqref{eq:boh30}, to find
\begin{equation*}%\label{eq:3000}
	\begin{aligned}
	 \frac{\mbox{RHS}_\e}{\e^{1-h}}=&-\int_{0}^T \varphi_3 \left( \int_{\omega_{12}}\e^{k-m}\partial_2\check{u}_{3\e} (x_1,x_2,0) \varphi_1  \varphi_2 \diff x_1 \diff x_2 \right)\diff x_3 + O(\e^{(1-h)/2})+O(\e^{M}).
	\end{aligned}		 
\end{equation*}

Since, by Lemma \ref{lemmatheta},  $\e^{k-m} \partial_2 \check{u}_{3\e}\weak \vartheta$ in $L^2(\check\Omega)$ from \eqref{eq:pluto} we deduce that $\e^{k-m} \partial_2 \check{u}_{3\e}\weak \vartheta$ also in $H^{1}((0,T),  W^{-1,2}(\omega_{12}))$. This in particular implies that
$$
	\begin{aligned}
\int_{\omega_{12}}\e^{k-m}\partial_2\check{u}_{3\e} (x_1,x_2,0) \varphi_1  \varphi_2 \diff x_1 \diff x_2&=\ <\varphi_1\otimes\varphi_2\otimes \delta_0,\e^{k-m}\partial_2\check{u}_{3\e}>
\\ 
&\to \
<\varphi_1\otimes\varphi_2\otimes \delta_0,\vartheta> =\int_{\omega_{12}}\vartheta (x_1) \varphi_1  \varphi_2 \diff x_1 \diff x_2
	\end{aligned}		
$$
from which we deduce that
\begin{equation}\label{eq:3000}
	\begin{aligned}
	 \frac{\mbox{RHS}_\e}{\e^{1-h}}\to-\int_{0}^T \varphi_3 \int_{\omega_{12}}\vartheta (x_1) \varphi_1  \varphi_2 \diff x_1 \diff x_2\diff x_3 
	 =\int_{\Omega_{J}} -\vartheta (x_1)\prod_{i=1}^{3} \varphi_i(x_i) \diff x_1 \diff x_2 
	 \diff x_3.	 
	 	\end{aligned}		 
\end{equation}

Rewriting \eqref{eq:boh30} as
$$
\frac{\e^{k+w}}{\e^m\e^{1-h}}
\frac{\e^m  \mbox{LHS}_\e}{\e^{k+w}}
=
 \frac{\mbox{RHS}_\e}{\e^{1-h}},
$$
noticing that $k+w-m+h-1=k+M-1$, and passing to the limit, by taking into account \eqref{eq:boh300}
and \eqref{eq:3000}, we conclude the proof.
\end{proof}

\subsection{Different Regimes for the Limit Junction Conditions}\label{sec:regimes}

\begin{figure}[!hbt]
	\centering
	\includegraphics[width=0.9\textwidth]{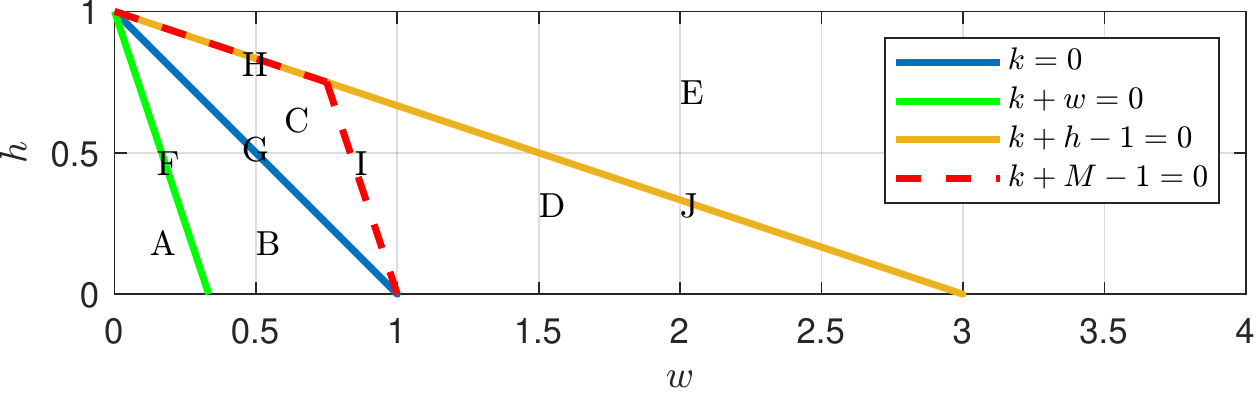}
	\caption{Junction conditions discriminant: displacements and torsion angle}
	\label{fig:cases}
\end{figure}

Considering \eqref{eq:jcond} and \eqref{eq:lastjc}, ten different cases are possible for the joining conditions, depending on the values of $w$ and $h$. The general scenario is graphically depicted in Fig.~\ref{fig:cases}, whereby the four lines represents the conditions $\hat{\varrho}_i=\check{\varrho}_i=1$ and $\hat{\varrho}_\vartheta=\check{\varrho}_\vartheta=1$. The possible cases have been labeled with letters from "A" to "J".

Analyzing Fig.~\ref{fig:cases}, it is noteworthy that for no combination of $w$ and $h$ (recall that $w$ and $h$ are strictly positive) the four joining conditions in (\ref{eq:jcond}) and \eqref{eq:lastjc} are non-trivial at the same time. The fictitious intersection, where all joining conditions would  be non-trivial at once, is at point $(w=0, h=1)$.  This corresponds to a non scaling of the stiffener in direction $x_2$ and to scaling with the same velocity the plate thickness and the stiffener dimension along $x_3$. This scaling leads the stiffener to degenerate into a prismatic portion of the plate: thus, the problem is equivalent to the asymptotic scaling of only a  plate.
\FloatBarrier

\section{The Limit Stored Energy}\label{sec:limitenergy}
In this Section, we characterize the limit stored energy, and we will prove our main $\Gamma$-convergence result.\\
To begin, in the next two Lemmata we characterize some components of the limit strain.

\begin{myLemma}\label{lemma:eplate}
	With the assumptions and notations of Lemma \ref{compattezzapiastra}, we have, up to subsequences, in the weak $L^2(\hat{\Omega})$ topology, 
	\begin{equation}
	\left(\hat{\mathbf{E}}_\e \hat{\mathbf{u}}_\e\right)_{\alpha\beta} \weak \frac{\partial_\alpha \hat{u}_\beta + \partial_\beta \hat{u}_\alpha}{2}
	\end{equation}
\end{myLemma}
\begin{proof}
It is sufficient to notice that $\left(\hat{\mathbf{E}}_\e \hat{\mathbf{u}}_\e\right)_{\alpha\beta} = \frac{\partial_\alpha \hat{u}_{\beta \e} + \partial_\beta \hat{u}_{\alpha \e}}{2}$ and to apply Lemma \ref{compattezzapiastra}.
\end{proof}

\begin{myLemma}\label{otherscond}
	With the assumptions and notation of Lemma \ref{lemmatheta} we have, up to subsequences, in the weak $L^2(\check{\Omega})$ topology:
	\begin{equation}\label{other1}
		\e^k \left(\check{\mathbf{E}}_\e \check{\mathbf{u}}_\e\right)_{11} \weak \partial_1 \check{u}_1,
	\end{equation}
	\begin{equation}\label{other2}
		\e^k \left(\check{\mathbf{E}}_\e \check{\mathbf{u}}_\e\right)_{13} \weak 
		\begin{cases}
			\frac{1}{2}\left(\partial_3\Phi + x_2\right)\partial_1\vartheta  & \textrm{ if } M=w=h,\\
			x_2 \partial_1\vartheta+\eta_{13} & \textrm{ if } M=w\neq h,\\
			0 & \textrm{ otherwise },
		\end{cases}
	\end{equation}
	\begin{equation}\label{other3}
		\e^k \left(\check{\mathbf{E}}_\e \check{\mathbf{u}}_\e\right)_{12} \weak  
		\begin{cases}
			\frac{1}{2}\left(\partial_2\Phi - (x_3 - \frac{H}{2})\right)\partial_1\vartheta & \textrm{ if } M=w=h,\\
			-(x_3 - \frac{H}{2}) \partial_1\vartheta+\eta_{12} & \textrm{ if } M=h\neq w,\\
			0 & \textrm{ otherwise },
		\end{cases}
	\end{equation}
	where $\eta_{13}\in L^2(\check{\Omega})$ is independent of $x_2$, $\eta_{12}\in L^2(\check{\Omega})$ is independent of $x_3$, and $\Phi\in L^2((-L,L), W^{1,2}(\check{\omega}))$ is the so-called torsion function, solution of the following boundary value problem:
	\begin{equation}\label{eq:laplaciantf}
	\begin{cases}
		\Delta \Phi = 0, & \text{ in } \check{\omega},\\
		\nabla \Phi \cdot \textbf{n} = -x_2 n_3 + (x_3-\frac{H}{2})n_2,   & \text{ on } \partial\check{\omega},\\
		\Rint_{\check{\omega}} \Phi \diff x_2\diff x_3 = 0,
	\end{cases}
\end{equation}
where $\textbf{n}$ is the outer normal to $\partial\check{\omega}$ and $\Delta(\cdot)$ is the Laplacian operator.
\end{myLemma}
\begin{proof}
	To prove (\ref{other1}), it is sufficient to notice that $\e^k \left(\check{\mathbf{E}}_\e \check{\mathbf{u}}_\e\right)_{11} = \e^k \partial_1 \check{u}_{1\e}$ and to apply Lemma \ref{compattezzatrave}.
	
	We have already deduced that, up to subsequences, 
	$\e^k \left(\check{\mathbf{E}}_\e \check{\mathbf{u}}_\e\right)_{1a} \weak \check{E}_{1a}$  in $L^2(\check{\Omega})$. 
	To characterize $\check{E}_{1a}$, note that
	\begin{equation*}
		\begin{split}
			2\partial_1\left( \check{\mathbf{W}}_\e \check{\mathbf{u}}_\e\right)_{23} &= \partial_1\left(\frac{\partial_3 \check{u}_{2\e}}{\e^{w+h}}-\frac{\partial_2 \check{u}_{3\e}}{\e^{w+h}}\right)\\
			&=\partial_3\left(\frac{\partial_1 \check{u}_{2\e}+\partial_2 \check{u}_{1\e}}{\e^{w+h}} \right)-\partial_2\left(\frac{\partial_1 \check{u}_{3\e}+\partial_3 \check{u}_{\e\,1}}{\e^{w+h}} \right)\\
			&=2\partial_3 \frac{\left( \check{\mathbf{E}}_\e \check{\mathbf{u}}_\e\right)_{12}}{\e^h}-2\partial_2 \frac{\left( \check{\mathbf{E}}_\e \check{\mathbf{u}}_\e\right)_{13}}{\e^w}
		\end{split}
	\end{equation*}
	in the sense of distributions. Hence, for $\psi\in C_0^\infty(\check{\Omega})$, we have
	\begin{equation}\label{eq:ciao}
		\int_{\check{\Omega}} \e^{k+M}\left(\check{\mathbf{W}}_\e \check{\mathbf{u}}_\e \right)_{23} \partial_1\psi \diff x= \int_{\check{\Omega}} \e^{k+M-h}\left(\check{\mathbf{E}}_\e \check{\mathbf{u}}_\e\right)_{12} \partial_3\psi \diff x-\int_{\check{\Omega}} \e^{k+M-w}\left(\check{\mathbf{E}}_\e \check{\mathbf{u}}_\e\right)_{13} \partial_2\psi \diff x.
	\end{equation}
	We note that, in $L^2(\check{\Omega})$,
	\begin{equation*}
		\begin{aligned}
			\e^{k+M-h}\left(\check{\mathbf{E}}_\e \check{\mathbf{u}}_\e\right)_{12} &\weak \begin{cases}
				\check{E}_{12}, & M=h,\\
				0, &\text{otherwise},
			\end{cases}
			& \e^{k+M-w}\left(\check{\mathbf{E}}_\e \check{\mathbf{u}}_\e\right)_{13} &\weak \begin{cases}
				\check{E}_{13}, & M=w,\\
				0, &\text{otherwise}.
			\end{cases}
		\end{aligned}
	\end{equation*}
	Passing to the limit in \eqref{eq:ciao}, we find
	\begin{equation*}
		\int_{\check{\Omega}} -\vartheta \partial_1\psi \diff x= \begin{cases}
			\int_{\check{\Omega}} \check{E}_{12} \partial_3\psi \diff x
			-\int_{\check{\Omega}} \check{E}_{13} \partial_2\psi \diff x, & M=h=w,\\
			-\int_{\check{\Omega}} \check{E}_{13} \partial_2\psi \diff x, & M=w\neq h, \\
			\int_{\check{\Omega}} \check{E}_{12} \partial_3\psi \diff x, & M=h\neq w.
		\end{cases}
	\end{equation*}
	Thus,
	\begin{equation*}
		\partial_1\vartheta =
		\begin{cases}
			\partial_2 \check{E}_{13}-\partial_3 \check{E}_{12} &\textrm{ if } M=h=w,\\
			\partial_2 \check{E}_{13}  & \textrm{ if } M=w\neq h,\\
			-\partial_3 \check{E}_{12} & \textrm{ if } M=h\neq w,
		\end{cases}
	\end{equation*} 
	and
	\begin{equation*}
		\begin{aligned}
			\check{E}_{13} &= 
			\begin{cases}
				x_2 \partial_1\vartheta+\gamma_{13}(x_1, x_3), & \textrm{ if } M=w\neq h,\\
				0, & \textrm{ if } M=h\neq w,
			\end{cases}
			&\check{E}_{12} &= 
			\begin{cases}
				
				0, & \textrm{ if } M=w\neq h,\\
				-x_3 \partial_1\vartheta+\gamma_{12}(x_1, x_2), & \textrm{ if } M=h\neq w.
			\end{cases}
		\end{aligned}
	\end{equation*}
	To conclude this part of the proof, we set $\eta_{13}\coloneqq\gamma_{13}$ and $\eta_{12}\coloneqq\gamma_{12} - \frac{H}{2}\partial_1\vartheta$.
	
	For the case $M=h=w$, in \cite[Lemma 4.1]{Freddi2010} it is shown that there exists a function 
	$\Phi\in L^2((-L,L),  W^{1,2}(\check{\omega}))$ satisfying the boundary value problem \eqref{eq:laplaciantf} and such that it can be written $\check{E}_{13} = \frac{1}{2}(\partial_3\Phi + x_2)\partial_1\vartheta$ and $\check{E}_{12} = \frac{1}{2}(\partial_2\Phi - (x_3 - H/2))\partial_1\vartheta$. $\Phi$ is commonly known in Mechanics as torsion function.
\end{proof}

Since we characterized only some components of the limit strain, the others will be defined by minimization of the stored energy density (see \eqref{f}). In fact, the convergences stated in Lemmata \ref{compattezzapiastra} and \ref{compattezzatrave} determine only some components of the limit strain energy, whilst the others remain undetermined. The minimization problem will select them in such a way to render the energy as small as possible. For this reason, we define
\begin{equation}\label{f0}
	\begin{split}
		\hat{f}_0(\hat{E}_{\alpha\beta})&\coloneqq\min\limits_{A_{i3}}\left\lbrace  {f}(\mathbf{A}):\mathbf{A}\in\Sym, \ A_{\alpha\beta} = \hat{E}_{\alpha\beta} \right\rbrace ,\\
		\check{f}_0(\check{E}_{1i})&\coloneqq
		\min\limits_{A_{ab}}\left\lbrace {f}(\mathbf{A}): \mathbf{A}\in\Sym, \  A_{1i}= \check{E}_{1i} \right\rbrace.
	\end{split}
\end{equation}
A direct computation shows that with
\begin{equation}\label{eq:zzs}
	\begin{aligned}
		\hat{\pmb{Z}} &\coloneqq \hat{E}_{\alpha\beta} \mathbf{e}_\alpha \otimes \mathbf{e}_\beta-\frac{\nu}{1-\nu} (\hat{E}_{11}+\hat{E}_{22})\mathbf{e}_3\otimes \mathbf{e}_3,\\
		\check{\pmb{Z}} &\coloneqq \check{E}_{11}\left[\mathbf{e}_1\otimes \mathbf{e}_1-\nu(\mathbf{e}_2\otimes \mathbf{e}_2 + \mathbf{e}_3\otimes \mathbf{e}_3)\right] + \check{E}_{1a}\left[\mathbf{e}_1\otimes \mathbf{e}_a+\mathbf{e}_a\otimes \mathbf{e}_1 \right].
	\end{aligned}
\end{equation}
we have
\begin{equation}\label{eq:energybeam}
	\begin{aligned}
		{f}(\hat{\mathbf{Z}}) &= \hat{f}_0(\hat{E}_{\alpha\beta}) = \frac{\textsf{E}}{2(1-\nu^2)}\left(\hat{E}_{11}^2 + \hat{E}_{22}^2 + 2\nu\hat{E}_{11}\hat{E}_{22} + 2(1-\nu)\hat{E}_{12}^2 \right),\\
		{f}(\check{\mathbf{Z}}) &= \check{f}_0(\check{E}_{1i}) = \frac{\textsf{E}}{2}\check{E}_{11}^2 + 2\mu \left(\check{E}_{12}^2 + \check{E}^2_{13} \right),
	\end{aligned}
\end{equation}
where $\textsf{E}\coloneqq\frac{\mu(2\mu+3\lambda)}{\mu+\lambda}$ is the Young modulus and $\nu \coloneqq \frac{\lambda}{2(\lambda+\mu)}$ is the Poisson ratio.
In particular, for the stiffener, we have the following characterization of the stored-energy density:
\begin{equation}\label{eq:beamdensity}
	\check{f}_0(\check{E}_{1i}) =\begin{cases}
		\check{f}_0\left(\partial_1\check{u}_1, \frac{1}{2}\left(\partial_2\Phi - (x_3-\frac{H}{2})\right)\partial_1\vartheta, \frac{1}{2}\left(\partial_3\Phi + x_2\right)\partial_1\vartheta\right), & \text{ if } M=h=w,\\
		\check{f}_0(\partial_1\check{u}_1, 0, x_2\partial_1\vartheta + \eta_{13}), & \text{ if } M=w \neq h,\\
		\check{f}_0(\partial_1\check{u}_1, -(x_3-\frac{H}{2})\partial_1\vartheta + \eta_{12}, 0), & \text{ if } M=h \neq w.\\
	\end{cases}
\end{equation}

\begin{myRemark}
	The characterization \eqref{eq:beamdensity} of the stored-energy density for the stiffener, combined with the ten regimes provided by all the possible junction conditions (see Sec. \ref{sec:regimes} and Fig. \ref{fig:cases}), results into twenty-three different limit problems.
	In particular, we have nine cases (i.e., A, B, C, D, E, F, G, I, J) for $M=w\neq h$, seven cases (i.e., A, B, C, E, F, G, H) for $M=h\neq w$ and seven cases (i.e., A, B, C, E, F, G, H (or I)) for $M=h=w$.
\end{myRemark}

For future convenience, let us introduce the set of the triads of the limit admissible displacements
\begin{equation*}
	\begin{multlined}
		\mathcal{A}\coloneqq\{ (\hat{\mathbf{u}}, \check{\mathbf{u}},\vartheta) \in KL_0(\hat{\Omega}) \times BN_0(\check{\Omega}) \times W^{1,2}_0((-L,L))  \text { satisfying }  (\ref{eq:jcond}), \  \eqref{eq:lastjc}
		\},
	\end{multlined}
\end{equation*}
the set 
\begin{equation*}
	\begin{multlined}
		\mathcal{A}_\e\coloneqq\{ (\hat{\mathbf{u}}_\e, \e^k\check{\mathbf{u}}_\e) \in W_0^{1,2}(\hat{\Omega}, \mathbb{R}^3) \times W_0^{1,2}(\check{\Omega}, \mathbb{R}^3)  \text { satisfying }  (\ref{eq:cond})\},
	\end{multlined}
\end{equation*}
and the extension of the stored energy 
$\mathcal{W}(\hat{\mathbf{u}},\check{\mathbf{u}},\vartheta):KL_0(\hat{\Omega})\times BN_0(\check{\Omega})\times W^{1,2}_0((-L,L))\to [0,\infty)$ defined by
\begin{equation*}
	\mathcal{W}(\hat{\mathbf{u}},\check{\mathbf{u}},\vartheta)\coloneqq 
	\begin{cases} 
		\hat{\mathcal{W}}(\hat{\mathbf{u}})+\check{\mathcal{W}}(\check{\mathbf{u}},\vartheta), & \mbox{if } (\hat{\mathbf{u}},\check{\mathbf{u}},\vartheta)\in \mathcal{A}, \\ \infty, & \mbox{ otherwise },
	\end{cases}
\end{equation*}
where 
\[\hat{\mathcal{W}}(\hat{\mathbf{u}}) \coloneqq \int_{\hat{\Omega}} \hat{f}_0(\hat{E}_{\alpha\beta}) \diff x\] and \[\check{\mathcal{W}}(\check{\mathbf{u}}, \vartheta) \coloneqq \int_{\check{\Omega}} \check{f}_0(\check{E}_{1i}) \diff x.\]

\begin{myTheorem}\label{gammatheorem}
As $\e \downarrow 0$, the sequence of functionals $\mathcal{W}_\e \left(\hat{\mathbf{u}}_{\e}, \e^k\check{\mathbf{u}}_{\e} \right)$ $\Gamma$-converges to the functional $\mathcal{W}(\hat{\mathbf{u}}, \check{\mathbf{u}}, \vartheta)$ in the following sense:
	\begin{enumerate}[(a)]
		\item (Liminf inequality) for every sequence $\e_n\downarrow 0$ and for every sequence $\{ \hat{\mathbf{u}}_{\e_n}, \e^k_n\check{\mathbf{u}}_{\e_n} \}\subset \mathcal{A}_{\e_n}$ such that
		\begin{equation*}
			\begin{aligned}
				&\hat{\mathbf{u}}_{\e_n}\weak \hat{\mathbf{u}}  &\text{ in } W^{1,2}(\hat{\Omega}, \mathbb{R}^{3}),\\
				&\e^k_n \check{ \mathbf{u}}_{\e_n}\weak \check{\mathbf{u}}  &\text{ in } W^{1,2}(\check{\Omega}, \mathbb{R}^{3}),\\
				& \e^{k+M}_n\left(\check{\mathbf{W}}_{\e_n}\check{\mathbf{u}}_{\e_n}\right)_{32}\weak \vartheta  &\text{ in } L^2(\check{\Omega}),
			\end{aligned}
		\end{equation*}
		we have
		\begin{equation*}
			\liminf\limits_{n\rightarrow\infty} \mathcal{W}_{\e_n}(\hat{\mathbf{u}}_{\e_n}, \e^k_n \check{\mathbf{u}}_{\e_n})\geq \mathcal{W}(\hat{\mathbf{u}}, \check{\mathbf{u}}, \vartheta);
		\end{equation*}
		\item (Existence of a recovery sequence) 
		assume either
		\begin{enumerate}
		\item[i)] {\it subcase G}: $k=0$, $h+w=1$, $1/2<M=w<1$, and $0<m=h<1/2$;
		\item[ii)] {\it subcase A}: $k<0$, $k+w<0$, $k+h-1<0$,  $M=h$, and  $m=w$;
		\item[iii)] {\it case E}: $k>0$, $k+w>0$, $k+h-1>0$ and $k+M-1>0$.
		\end{enumerate}
		
		For every sequence $\e_n\downarrow 0$ and for every  $\left(\hat{\mathbf{u}}, \check{\mathbf{u}}, \vartheta \right)\in \mathcal{A}$ there exists a sequence $\{ \hat{\mathbf{u}}_{\e_n}, \e^k_n\check{\mathbf{u}}_{\e_n} \}\subset \mathcal{A}_{\e_n}$, called recovery sequence, such that
		\begin{equation*}
			\begin{aligned}
				&\hat{\mathbf{u}}_{\e_n}\weak \hat{\mathbf{u}}  &\text{ in } W^{1,2}(\hat{\Omega}, \mathbb{R}^{3}),\\
				&\e^k_n \check{ \mathbf{u}}_{\e_n}\weak \check{\mathbf{u}}  &\text{ in } W^{1,2}(\check{\Omega}, \mathbb{R}^{3}),\\
				& \e^{k+M}_n\left(\check{\mathbf{W}}_{\e_n}\check{\mathbf{u}}_{\e_n}\right)_{32}\weak \vartheta  &\text{ in } L^2(\check{\Omega}),
			\end{aligned}
		\end{equation*}
		and
		\begin{equation*}
			\lim\limits_{n\rightarrow\infty} \mathcal{W}_{\e_n}(\hat{\mathbf{u}}_{\e_n}, \e^k_n \check{\mathbf{u}}_{\e_n})= \mathcal{W}(\hat{\mathbf{u}}, \check{\mathbf{u}}, \vartheta).
		\end{equation*}
	\end{enumerate}
\end{myTheorem}

\begin{proof}
	~\paragraph{(a) Liminf inequality}
	We start by proving the weak sequential lower semicontinuity of the family of stored energy functionals. Without loss of generality, we can suppose that
	$\liminf\limits_{n\rightarrow\infty} \mathcal{W}_{\e_n}(\hat{\mathbf{u}}_{\e_n}, \e^k_n \check{\mathbf{u}}_{\e_n}) < \infty$
	otherwise there is nothing to prove. Hence,
	$\sup\limits_{n} \mathcal{W}_{\e_n}(\hat{\mathbf{u}}_{\e_n}, \e^k_n \check{\mathbf{u}}_{\e_n}) < \infty$, and Lemmata \ref{compattezzapiastra}, \ref{compattezzatrave}, \ref{lemmatheta}, \ref{lemma:eplate}, and \ref{otherscond} hold.
	Taking into account the decomposition given in (\ref{energy}), we need to show the weak sequential lower semicontinuity of the stored energy contributions due to the plate and the stiffener, i.e.
	\begin{equation*}
		\begin{aligned}
			\liminf\limits_{n\rightarrow\infty} \hat{\mathcal{W}}_{\e_n}(\hat{\mathbf{u}}_{\e_n})&\geq \int_{\hat{\Omega}} \hat{f}(\hat{\mathbf{Z}}) \diff x,
			&\liminf\limits_{n\rightarrow\infty} \check{\mathcal{W}}_{\e_n}(\e^k_n \check{\mathbf{u}}_{\e_n})&\geq \int_{\check{\Omega}} \check{f}(\check{\mathbf{Z}}) \diff x
		\end{aligned}
	\end{equation*}
for every sequence $\hat{\mathbf{u}}_{\e_n} \weak \hat{\mathbf{u}}$ in $W^{1,2}(\hat{\Omega}, \mathbb{R}^{3})$ and $\e_n^k\check{\mathbf{u}}_{\e_n} \weak \check{\mathbf{u}}$ in $W^{1,2}(\check{\Omega}, \mathbb{R}^{3})$, respectively.\\
It is easy to prove, by an application of Fatou Lemma, that $\hat{\mathcal{W}}_{\e_n}(\hat{\mathbf{u}}_{\e_n})$ is sequential lower semicontinuous with respect to the strong $W^{1,2}(\hat{\Omega}, \mathbb{R}^{3})$ topology. However, the convexity of the integrand function $\hat{f}(\cdot)$ is sufficient (yet not necessary in the vector-valued case) to ensure the sequential lower semicontinuity also with respect to the weak $W^{1,2}(\hat{\Omega}, \mathbb{R}^{3})$ topology (see, for instance, \cite[Proposition 1.18]{Maso1993}, \cite[Theorem 2.6]{Rindler2018}). By using  (\ref{f0}), we infer
\begin{equation*}
		\liminf\limits_{n\rightarrow \infty}\hat{\mathcal{W}}_{\e_n}(
		\hat{\mathbf{u}}_{\e_n}) = 
		\liminf\limits_{n\rightarrow \infty} \int_{\hat{\Omega}} \hat{\chi}_{\e_n}\hat{f}\left( \hat{\mathbf{E}}_{\e_n} \hat{\mathbf{u}}_{\e_n}\right) \diff x 
		\geq\int_{\hat{\Omega}}\hat{f}\left(\hat{\mathbf{E}}\right) \diff x
		\geq \int_{\hat{\Omega}}\hat{f} \left(\hat{\pmb{Z}}\right) \diff x \left(=: \int_{\hat{\Omega}}\hat{f}_0 \left(\hat{E}_{\alpha\beta}\right) \diff x\right).
\end{equation*}
Similarly, for the stiffener, we infer
	\begin{equation*}
			\liminf\limits_{n\rightarrow \infty}\check{\mathcal{W}}_{\e_n}(\e^k
			\check{\mathbf{u}}_{\e_n}) = \liminf\limits_{n\rightarrow \infty} \int_{\check{\Omega}} \check{\chi}_{\e_n} \check{f}\left( \e^k_n \check{\mathbf{E}}_{\e_n} \check{\mathbf{u}}_{\e_n}\right) \diff x 
			\geq \int_{\check{\Omega}}\check{f} \left(\check{\pmb{E}}\right) \diff x\geq \int_{\check{\Omega}}\check{f} \left(\check{\pmb{Z}}\right) \diff x \left(=: \int_{\check{\Omega}}\check{f}_0\left(\check{E}_{1i}\right) \diff x\right),
	\end{equation*}
and from Lemma \ref{otherscond}, we have
	\begin{equation*}
		\int_{\check{\Omega}}\check{f}_0\left(\check{E}_{1i}\right) \diff x = \begin{cases}
			\int_{\check{\Omega}} \check{f}_0\left(\partial_1\check{u}_1, \frac{1}{2}\left(\partial_2\Phi - (x_3-\frac{H}{2})\right)\partial_1\vartheta, \frac{1}{2}\left(\partial_3\Phi + x_2\right)\partial_1\vartheta\right) \diff x, & \text{ if } M=h=w,\\
			\int_{\check{\Omega}} \check{f}_0(\partial_1\check{u}_1, 0, x_2\partial_1\vartheta + \eta_{13}) \diff x, & \text{ if } M=w \neq h,\\
			\int_{\check{\Omega}} \check{f}_0(\partial_1\check{u}_1, -(x_3-\frac{H}{2})\partial_1\vartheta + \eta_{12}, 0) \diff x, & \text{ if } M=h \neq w.
		\end{cases}
	\end{equation*}
	A direct computation shows that, for the latter two cases, the estimation is independent from $\eta_{12}$ and $\eta_{13}$. We provide the computation for the case $M=w\neq h$ only, the other one being conceptually similar.
	We have 
	\begin{equation*}
		\begin{aligned}
			\int_{\check{\Omega}} \check{f}_0(\partial_1\check{u}_1, 0, x_2\partial_1\vartheta + \eta_{13}) \diff x &= \begin{multlined}[t]
				\int_{\check{\Omega}} \check{f}_0(\partial_1\check{u}_1, 0, x_2\partial_1\vartheta) \diff x + \\ +2\mu \int_{\check{\Omega}} \eta_{13}^2 \diff x  + 4\mu \int_{\check{\Omega}} x_2 \partial_1 \vartheta \eta_{13} \diff x \end{multlined}\\
			&\geq \int_{\check{\Omega}} \check{f}_0(\partial_1\check{u}_1, 0, x_2\partial_1\vartheta) \diff x,
		\end{aligned}
	\end{equation*}
	since the integral of $x_2 \partial_1 \vartheta \eta_{13}$ is zero because $\eta_{13}$ does not depend on $x_2$ by Lemma \ref{otherscond}. Consequently, we have that 
	\begin{equation}
		\liminf\limits_{n\rightarrow \infty}\check{\mathcal{W}}_{\e_n}(\e^k
		\check{\mathbf{u}}_{\e_n}) \geq \begin{cases}
			\int_{\check{\Omega}} \check{f}_0\left(\partial_1\check{u}_1, \frac{1}{2}\left(\partial_2\Phi - (x_3-\frac{H}{2})\right)\partial_1\vartheta, \frac{1}{2}\left(\partial_3\Phi + x_2\right)\partial_1\vartheta\right) \diff x, & \text{ if } M=h=w,\\
			\int_{\check{\Omega}} \check{f}_0(\partial_1\check{u}_1, 0, x_2\partial_1\vartheta) \diff x, & \text{ if } M=w \neq h,\\
			\int_{\check{\Omega}} \check{f}_0(\partial_1\check{u}_1, -(x_3-\frac{H}{2})\partial_1\vartheta, 0) \diff x, & \text{ if } M=h \neq w.
		\end{cases}
	\end{equation}

	~\paragraph{(b) Existence of a recovery sequence}
	~\subparagraph{case \textit{i})}
	We start by proving case {\it i).} So, let $k=0$, $h+w=1$, $1/2<M=w<1$, and $0<m=h<1/2$.
	
		Let $(\hat{\mathbf{u}},\check{\mathbf{u}},\vartheta)\in\mathcal{A}$. From the definition of $\mathcal{A}$ we have that
$$
\hat{u}_{1} = \hat{\xi}_1(x_1, x_2) - x_3\partial_1 \hat{\xi}_3(x_1, x_2),
\quad
\hat{u}_{2} = \hat{\xi}_2(x_1, x_2) - x_3\partial_2 \hat{\xi}_3(x_1, x_2),
\quad
\hat{u}_{3} = \hat{\xi}_3(x_1, x_2),
$$
and
$$
\check u_1 = \check{\xi}_1(x_1)-x_2\partial_1\check{\xi}_2(x_1)-x_3\partial_1\check{\xi}_3(x_1), \quad
\check u_2 = \check{\xi}_2(x_1), \quad \check u_3 = \check{\xi}_3(x_1),
$$	
for appropriate functions $\hat{\pmb{\xi}}$ and $\check{\pmb{\xi}}$.
With the values in consideration of $k, h$ and $w$, Lemmata \ref{lemma:limitjoind} and \ref{lemma:limitjoint} imply that
\begin{equation}\label{lemmataboh}
\hat \xi_1(x_1,0)=\check \xi_1(x_1)\quad \check \xi_2(x_1)=0, \quad \hat \xi_3(x_1,0)=0, \quad \partial_2\hat \xi_3(x_1,0)=0.
\end{equation}

To start, we assume
$\hat{\xi}_i \in C^\infty((-L,L)\times(-L,L))$, and
 $\check{\xi}_3, \vartheta\in C^\infty((-L,L))$.
 Moreover, we assume that all these functions have value zero in a neighborhood of $x_1=L$ and, in view of \eqref{lemmataboh}, we may
also assume that $\hat{\xi}_3$ is equal to zero in a neighborhood of $(-L,L)\times\{0\}$. 
 
Let the sequences $\hat{\mathbf{u}}^{\flat}_\e$ and $\check{\mathbf{u}}^{\flat}_\e$ be defined by:
	\begin{equation*}
		\begin{aligned}
			\hat{u}^{\flat}_{1\e} &\coloneqq - x_3\partial_1\hat{\xi}_3(x_1,x_2)
			-\e^w x_3 \partial_1 \check{\xi}_3(x_1) + \hat{r}^\flat_{1\e},\\
			\hat{u}^{\flat}_{2\e} &\coloneqq  - x_3\partial_2\hat{\xi}_3(x_1,x_2)
			-\e^h x_3 \vartheta(x_1) + \hat{r}^\flat_{2\e},\\
			\hat{u}^{\flat}_{3\e} &\coloneqq \hat{\xi}_3(x_1, x_2)+
			\e^h x_2 \vartheta(x_1) + \e^w \check{\xi}_3(x_1) + \hat{r}^\flat_{3\e},
		\end{aligned}
	\end{equation*}
	and					
	\begin{equation*}
		\begin{aligned}
			\check{u}^{\flat}_{1\e} &\coloneqq
			\hat{\xi}_1(x_1,0)-x_3 \partial_1\check{\xi}_3(x_1)-\e^w x_2\partial_1\hat{\xi}_2(x_1,0) + \e^h x_2 x_3 \partial_1 \vartheta(x_1),\\
			\check{u}^{\flat}_{2\e} &\coloneqq
			\e^w\hat{\xi}_2(x_1,0) - \e^h x_3 \vartheta(x_1) -\nu \e^{2w}x_2\partial_1\hat{\xi}_1(x_1,0) + \nu \e^{2w}x_2x_3\check\psi(x_3)\partial^2_1\check{\xi}_3(x_1), \\
			\check{u}^{\flat}_{3\e} &\coloneqq
			\check{\xi}_3(x_1) + \e^h x_2 \vartheta(x_1) -\nu \e^{2h}x_3\check\psi(x_3) \left[ - \frac{x_3}{2}\partial^2_1\check{\xi}_3(x_1) + \partial_1\hat{\xi}_1(x_1, 0)\right],
		\end{aligned}
	\end{equation*}
	with
	\begin{equation*}
		\begin{aligned}
			\hat{r}^\flat_{1\e} (x_1, x_2, x_3) &\coloneqq -(1-\hat\psi)x_2 \partial_1\hat{\xi}_2(x_1,0) + (1-2\hat\psi)\e^h x_2 x_3 \partial_1 \vartheta(x_1)+\begin{cases}
				\hat{\xi}_1(x_1,0), & \text{ if } | x_2| \leq \e^w W,\\
				\hat{\xi}_1(x_1,2(|x_2|-\e^w W)),  & \text{ if } \e^w W\leq | x_2| \leq 2\e^w W,\\
				\hat{\xi}_1(x_1,x_2), & \text{ if } | x_2| \geq 2\e^w W,\\
			\end{cases}\\
			\hat{r}^\flat_{2\e} (x_1, x_2) &\coloneqq -(1-\hat\psi)\nu x_2\partial_1\hat{\xi}_{1}(x_1,0)+\begin{cases}
				\hat{\xi}_2(x_1,0), & \text{ if } | x_2| \leq \e^w W,\\
				\hat{\xi}_2(x_1,2(|x_2|-\e^w W)), & \text{ if } \e^w W\leq | x_2| \leq 2\e^w W,\\
				\hat{\xi}_2(x_1,x_2) , & \text{ if } | x_2| \geq 2\e^w W,\\
			\end{cases}\\
			\hat{r}^\flat_{3\e} (x_1, x_2, x_3) &\coloneqq -\hat\psi \e^2 \frac{\nu}{1-\nu}\left[x_3 \left(\partial_1\hat{\xi}_1(x_1, x_2) + \partial_2\hat{\xi}_2(x_1, x_2)\right) - \frac{x_3^2}{2}\left(\partial^2_1\hat{\xi}_3(x_1, x_2) + \partial^2_2\hat{\xi}_3(x_1, x_2) \right)\right],\\
		\end{aligned}
	\end{equation*}
where
$\check \psi(x_3):[0,H]\to[0,1]$ is the continuous piece-wise affine function taking value equal to $0$ in $[0,T\e^{1-h}]$,
$1$ in $[2T\e^{1-h}, H]$, and being affine in $[T\e^{1-h},2T\e^{1-h}]$, whilst $\hat \psi(x_2):[-L, L]\to[0,1]$ is the continuous piece-wise affine function taking value equal to $0$ in $[-\e^w W, \e^w W]$,
$1$ in $[-L, -2\e^w W] \cup [2\e^w W, L]$, and being affine in $[\e^w W, 2\e^w W]$ and $[-2\e^w W, -\e^w W]$.

Note that the sequences $\hat{\mathbf{u}}^{\flat}_\e$, $\check{\mathbf{u}}^{\flat}_\e$ are continuous at the interfaces, i.e., at $x_2=\pm \e^w W$ and $x_2=\pm 2\e^w W$. Due to the smoothness of $\hat{\pmb{\xi}}$, $\check{\pmb{\xi}}$, and $\vartheta$, we can conclude that $\hat{\mathbf{u}}^{\flat}_\e$, $\check{\mathbf{u}}^{\flat}_\e$ are (at least) of class $W^{1,2}$.
	It can be easily verified that the pair $(\hat{\mathbf{u}}^{\flat}_\e, \check{\mathbf{u}}^{\flat}_\e)$ satisfies the boundary conditions at $x_1 = L$ and the junction conditions \eqref{eq:cond} for $\e$ small enough, since $\hat{\xi}_3$ is equal to zero in a neighborhood of $(-L,L)\times\{0\}$.

Since  $|\partial_2\hat\psi(x_2)|\lesssim \e^{-w}$,  $|\partial_2(x_2\hat\psi(x_2))|\le 3$ for every $x_2$, and since $|\partial_3\check\psi(x_3)|\lesssim \e^{h-1}$,  $|\partial_3(x_3\check\psi(x_3))|\le 3$ for every $x_3$, it follows that
$$
\hat{\mathbf{u}}^{\flat}_\e \to \hat{\mathbf{u}} \quad \text{ in } W^{1,2}(\hat{\Omega}, \mathbb{R}^{3}), 
\qquad
\e^k 
\check{\mathbf{u}}^{\flat}_\e \to\check{\mathbf{u}} \quad\text{ in } W^{1,2}(\check{\Omega}, \mathbb{R}^{3}).
$$
It is easily checked that
$$
\e^k \check{\mathbf{E}}_\e \check{\mathbf{u}}^{\flat}_\e \to \check{\pmb{Z}} \quad \text{ in } L^2(\check{\Omega}, \mathbb{R}^{3\times 3}),
\qquad
\e^{k+M} \left(\check{\mathbf{W}}_\e \check{\mathbf{u}}^{\flat}_\e\right)_{32} \to\vartheta \quad \text{ in } L^2(\check{\Omega}), \\
$$
where (see Eq. \eqref{eq:zzs})
$$
\check{\pmb{Z}} \coloneqq \partial_1 \check{u}_{1}\left[\mathbf{e}_1\otimes \mathbf{e}_1-\nu(\mathbf{e}_2\otimes \mathbf{e}_2 + \mathbf{e}_3\otimes \mathbf{e}_3)\right] +x_2\partial_1\vartheta\left[\mathbf{e}_1\otimes \mathbf{e}_3+\mathbf{e}_3\otimes \mathbf{e}_1 \right].
$$
Let now (see Eq. \eqref{eq:zzs})
\begin{equation*}
		\hat{\pmb{Z}} \coloneqq  \frac{\partial_\alpha\hat{u}_{\beta}+\partial_\beta\hat{u}_{\alpha}}2 \mathbf{e}_\alpha \otimes \mathbf{e}_\beta-\frac{\nu}{1-\nu} (\partial_1\hat{u}_{1}+\partial_2\hat{u}_{2})\mathbf{e}_3\otimes \mathbf{e}_3.
\end{equation*}
From the convergence $\hat{\mathbf{u}}^{\flat}_\e \to \hat{\mathbf{u}}$ in $W^{1,2}(\hat{\Omega}, \mathbb{R}^{3})$ it immediately follows that $(\hat{\mathbf{E}}_\e \hat{\mathbf{u}}^{\flat}_\e )_{\alpha\beta}\to (\hat{\pmb{Z}})_{\alpha\beta}$ in $L^{2}(\hat{\Omega})$.
A short computation shows that the component $13$ is
$$
(\hat{\mathbf{E}}_\e \hat{\mathbf{u}}^{\flat}_\e )_{13}  = 
\begin{cases}
\displaystyle	\frac{\e^h x_2 \partial_1\vartheta}\e, & \text{ if } | x_2| \leq \e^w W,\\
\displaystyle	[1+\partial_3\left(x_3(1- 2\hat\psi)\right)]\frac{\e^h x_2 \partial_1\vartheta}\e, & \text{ if } \e^w W\leq | x_2| \leq 2\e^w W,\\
	 0, & \text{ if } | x_2| \geq 2\e^w W,\\
\end{cases}
$$
and recalling that $h+w=1$ we deduce that
$$
|(\hat{\mathbf{E}}_\e \hat{\mathbf{u}}^{\flat}_\e )_{13}|  \lesssim 
\begin{cases}
\displaystyle	|\partial_1\vartheta|, & \text{ if } | x_2| \leq 2\e^w W,\\
	 0, & \text{ if } | x_2| \geq 2\e^w W,\\
\end{cases}
$$	 
from which it follows that $(\hat{\mathbf{E}}_\e \hat{\mathbf{u}}^{\flat}_\e )_{13}\to 0$ in $L^{2}(\hat{\Omega})$.	
With similar arguments we arrive at
$$
\hat{\mathbf{E}}_\e \hat{\mathbf{u}}^{\flat}_\e \to \hat{\pmb{Z}} \quad \text{ in } L^2(\hat{\Omega}, \mathbb{R}^{3\times 3}).
$$

The strong convergence of the rescaled strains leads to
\begin{eqnarray*}
\lim_{\e\rightarrow 0} \mathcal{W}_{\e}(\hat{\mathbf{u}}_{\e}^{\flat}, \e^k \check{\mathbf{u}}_{\e}^{\flat})
& = &
\displaystyle
\lim_{\e\rightarrow 0} 
\frac{1}{2}\int_{\hat{\Omega}}\hat{\chi}_\e\mathbb{C}\left[\hat{\mathbf{E}}_\e\hat{\mathbf{u}}_\e^{\flat}\right]\cdot \hat{\mathbf{E}}_\e\hat{\mathbf{u}}_\e^{\flat} \diff x  +\frac{1}{2}\int_{\check{\Omega}} \check{\chi}_\e\mathbb{C}\left[\e^{k}\check{\mathbf{E}}_\e\check{\mathbf{u}}_\e^{\flat}\right]\cdot \e^{k}\check{\mathbf{E}}_\e\check{\mathbf{u}}_\e^{\flat} \diff x\\
& = &
\displaystyle
\frac{1}{2}\int_{\hat{\Omega}}\mathbb{C}\left[\hat{\pmb{Z}}\right]\cdot \hat{\pmb{Z}} \diff x  +
\frac{1}{2}\int_{\check{\Omega}} \mathbb{C}\left[\check{\pmb{Z}}\right]\cdot \check{\pmb{Z}} \diff x
=\int_{\hat{\Omega}} f(\hat{\pmb{Z}}) \diff x+\int_{\check{\Omega}} f(\check{\pmb{Z}}) \diff x
\\
& =  &
\int_{\hat{\Omega}} \hat f_0(\hat{{Z}}_{\alpha\beta}) \diff x+\int_{\check{\Omega}} \check f_0(\check{{Z}}_{1i}) \diff x
=\mathcal{W}(\hat{\mathbf{u}}, \check{\mathbf{u}}, \vartheta),
\end{eqnarray*}
where we have used the definition \eqref{f0} and the identity \eqref{eq:energybeam}.

Thus, under the regularity assumptions made, $\{\hat{\mathbf{u}}^{\flat}_{\e_n}, \check{\mathbf{u}}^{\flat}_{\e_n}\}$ is a recovery sequence.
	The general case $\{\hat{\mathbf{u}}^{\flat}_{\e_n}, \check{\mathbf{u}}^{\flat}_{\e_n}\}$ is achieved by approximating $(\hat{\mathbf{u}},\check{\mathbf{u}},\vartheta)$ in $\mathcal{A}$ by a sequence from $\mathcal{A} \cap\left(C^\infty(\hat{\Omega}, \mathbb{R}^{3}) \times C^\infty(\check{\Omega}, \mathbb{R}^{3}) \times C^\infty((-L,L)) \right)$ that strongly converges to $(\hat{\mathbf{u}},\check{\mathbf{u}},\vartheta)$ in $W^{1,2}(\hat{\Omega}, \mathbb{R}^3)\times W^{1,2}(\check{\Omega}, \mathbb{R}^3)\times W^{1,2}((-L,L))$, concluding with a standard diagonal argument.

\vspace{3mm}
~\subparagraph{case \textit{ii})}
We now prove case {\it ii)}.
Let $k<0$, $k+w<0$, $k+h-1<0$,  $M=h$, and $m=w$. 
	Let $(\hat{\mathbf{u}},\check{\mathbf{u}},\vartheta)\in\mathcal{A}$. From the definition of $\mathcal{A}$ we have that
$$
\hat{u}_{1} = \hat{\xi}_1(x_1, x_2) - x_3\partial_1 \hat{\xi}_3(x_1, x_2),
\quad
\hat{u}_{2} = \hat{\xi}_2(x_1, x_2) - x_3\partial_2 \hat{\xi}_3(x_1, x_2),
\quad
\hat{u}_{3} = \hat{\xi}_3(x_1, x_2),
$$
and
$$
\check u_1 = \check{\xi}_1(x_1)-x_2\partial_1\check{\xi}_2(x_1)-x_3\partial_1\check{\xi}_3(x_1), \quad
\check u_2 = \check{\xi}_2(x_1), \quad \check u_3 = \check{\xi}_3(x_1),
$$	
for appropriate functions $\hat{\pmb{\xi}}$ and $\check{\pmb{\xi}}$.
In the case under study, Lemmata \ref{lemma:limitjoind} and \ref{lemma:limitjoint} state that
\begin{equation}\label{lemmataA}
\hat \xi_1(x_1,0)=\hat \xi_2(x_1,0)=\hat \xi_3(x_1,0)=0, \quad\partial_2\hat \xi_3(x_1,0)=0.
\end{equation}

We first build the recovery sequence assuming that 	
$\hat{\xi}_i \in C^\infty((-L,L)\times(-L,L))$, and
 $\check{\xi}_i, \vartheta\in C^\infty(-L,L)$.
 Moreover, we assume that all these functions have value zero in a neighborhood of $x_1=L$, and in view of \eqref{lemmataA},
 that $\hat{\xi}_i$ are equal to zero in a neighborhood of $(-L,L)\times\{0\}$. 
 
To define the recovery sequence for the stiffener, $\check{\mathbf{u}}^{\flat}_\e$, we shall use the continuous piece-wise affine function  $\check \psi(x_3):[0,H]\to[0,1]$  defined as in the previous case. We set
\begin{equation}\label{eq:recAcheck}
	\begin{aligned}
		\e^k\check{u}_{1\e}^\flat&\coloneqq
		\check{\xi}_1(x_1)-x_2\partial_1\check{\xi}_2(x_1)-x_3\partial_1\check{\xi}_3(x_1) - \e^{w} x_2 x_3\partial_1 \vartheta(x_1)
		+\nu \e^{2w}\frac{x_2^2}2x_3\partial_1^3\check{\xi}_3(x_1),\\
		\e^k\check{u}_{2\e}^\flat&\coloneqq
		\check{\xi}_2(x_1)- \e^{w} x_3 \vartheta(x_1)-\nu \e^{2w}\Big(x_2\partial_1\check{\xi}_1(x_1)-\frac{x_2^2}2\partial_1^2\check{\xi}_2(x_1)-x_2x_3\partial_1^2\check{\xi}_3(x_1)\Big),
		\\	
		\e^k\check{u}_{3\e}^\flat&\coloneqq
		\check{\xi}_3(x_1)+\e^{w} x_2 \vartheta(x_1)-\nu \e^{2h}\check\psi(x_3)\Big(x_3\partial_1\check{\xi}_1(x_1)-x_2x_3\partial_1^2\check{\xi}_2(x_1)-\frac{x_3^2}2\partial_1^2\check{\xi}_3(x_1)\Big)-\nu \e^{2w}\frac{x_2^2}2\partial_1^2\check{\xi}_3(x_1).	
	\end{aligned}
\end{equation}
Since $|\partial_3(x_3\check\psi(x_3)) |\le 3$ for every $x_3$, it follows that
$$
\check{\mathbf{u}}^{\flat}_\e \to\check{\mathbf{u}} \quad \text{ in } W^{1,2}(\check{\Omega}, \mathbb{R}^{3}),
\quad \mbox{and}\quad
\e^{k+M} \left(\check{\mathbf{W}}_\e \check{\mathbf{u}}^{\flat}_\e\right)_{32} \to \vartheta  \text{ in } L^2(\check{\Omega}),
$$
and
$$
		\e^k \check{\mathbf{E}}_\e \check{\mathbf{u}}^{\flat}_\e  \to \check{\pmb{Z}}  \quad \text{ in } L^2(\check{\Omega}, \mathbb{R}^{3\times 3}),
$$
where (see Eq. \eqref{eq:zzs})
$$
\check{\pmb{Z}} \coloneqq \partial_1 \check{u}_{1}\left[\mathbf{e}_1\otimes \mathbf{e}_1-\nu(\mathbf{e}_2\otimes \mathbf{e}_2 + \mathbf{e}_3\otimes \mathbf{e}_3)\right] -x_3\partial_1\vartheta\left[\mathbf{e}_1\otimes \mathbf{e}_2+\mathbf{e}_2\otimes \mathbf{e}_1 \right].
$$
The recovery sequence for the plate  $\hat{\mathbf{u}}^{\flat}_\e$ is defined by

\begin{equation}\label{eq:recAhat}
	\begin{aligned}
		\hat{u}_{1\e}^\flat &\coloneqq \hat{\xi}_1(x_1, x_2) - x_3\partial_1 \hat{\xi}_3(x_1, x_2)+\check{u}_{1\e}^\flat(x_1,\e^{-w}x_2,\e^{1-h}x_3),\\
		\hat{u}_{2\e}^\flat &\coloneqq \hat{\xi}_2(x_1, x_2) - x_3\partial_2 \hat{\xi}_3(x_1, x_2)+\e^{-w}\check{u}_{2\e}^\flat(x_1,\e^{-w}x_2,\e^{1-h}x_3),\\
		\hat{u}_{3\e}^\flat &\coloneqq \hat{\xi}_3(x_1, x_2) - \e^2 \frac{\nu}{1-\nu}\left[x_3\left(\partial_1 \hat{\xi}_1(x_1, x_2)+ \partial_2 \hat{\xi}_2(x_1, x_2)\right) - \frac{x_3^2}{2}\left(\partial^2_1 \hat{\xi}_3(x_1, x_2) + \partial^2_2 \hat{\xi}_3(x_1, x_2)\right) 
		\right]\\
		&\quad+\e^{1-h}\check{u}_{3\e}^\flat(x_1,\e^{-w}x_2,\e^{1-h}x_3).
			\end{aligned}
\end{equation}
Since $\hat{\xi}_i$ are equal to zero in a neighborhood of $(-L,L)\times\{0\}$, for $\e$ small enough the junction conditions \eqref{eq:cond} are automatically satisfied.

For $x_3\in[0,T]$ we have that

\begin{equation*}
	\begin{aligned}
\check{u}_{1\e}^\flat(x_1,\e^{-w}x_2,\e^{1-h}x_3)&=\e^{-k}\big[
\check{\xi}_1-\e^{-w}x_2\partial_1\check{\xi}_2-\e^{1-h}x_3\partial_1\check{\xi}_3 - \e^{1-h} x_2 x_3\partial_1 \vartheta+\nu \frac{x_2^2}2x_3\partial_1^3\check{\xi}_3(x_1)\big]\\
\e^{-w}\check{u}_{2\e}^\flat(x_1,\e^{-w}x_2,\e^{1-h}x_3)&=\e^{-k-w}\big[
\check{\xi}_2- \e^{w+1-h} x_3 \vartheta-\nu (\e^{w}x_2\partial_1\check{\xi}_1-\frac{x_2^2}2\partial_1^2\check{\xi}_2-\e^{w+1-h}x_2x_3\partial_1^2\check{\xi}_3)
\big]\\
\e^{1-h}\check{u}_{3\e}^\flat(x_1,\e^{-w}x_2,\e^{1-h}x_3)&=
\e^{1-h-k}\big[
\check{\xi}_3+ x_2 \vartheta-\nu \frac{x_2^2}2\partial_1^2\check{\xi}_3
\big]
\end{aligned}
\end{equation*}
and since $-k,1-h-k$, and $-k-w$ are strictly greater than zero, it follows that
$$
\hat{\mathbf{u}}^{\flat}_\e \to\check{\mathbf{u}} \quad \text{ in } W^{1,2}(\hat{\Omega}, \mathbb{R}^{3}).
$$
A tedious calculation then shows that
$$
\hat{\mathbf{E}}_\e \hat{\mathbf{u}}^{\flat}_\e \to \hat{\pmb{Z}} \quad \text{ in } L^2(\hat{\Omega}, \mathbb{R}^{3\times 3}).
$$
where (see Eq. \eqref{eq:zzs})
\begin{equation}\label{hatZr}
		\hat{\pmb{Z}} \coloneqq  \frac{\partial_\alpha\hat{u}_{\beta}+\partial_\beta\hat{u}_{\alpha}}2 \mathbf{e}_\alpha \otimes \mathbf{e}_\beta-\frac{\nu}{1-\nu} (\partial_1\hat{u}_{1}+\partial_2\hat{u}_{2})\mathbf{e}_3\otimes \mathbf{e}_3.
\end{equation}
The proof is concluded as in case {\it i)}

\vspace{3mm}
~\subparagraph{case \textit{iii})}
We finally prove case {\it iii)}.
Let $k>0$, $k+w>0$, $k+h-1>0$ and $k+M-1>0$. In this case, from Lemmata \ref{lemma:limitjoind} and \ref{lemma:limitjoint} we deduce that $\check{\xi}_i=0$ and $\vartheta=0$. As a consequence, the limit displacement $\hat{\textbf{u}}$ as well as the limit strain tensor $\check{\textbf{E}}$ (and $\check{\textbf{Z}}$) of the beam are identically equal to zero.

To start, we assume $(\hat{\mathbf{u}},\textbf{0},0)\in\mathcal{A} \cap\left(C^\infty(\hat{\Omega}, \mathbb{R}^{3}) \times C^\infty(\check{\Omega}, \mathbb{R}^{3}) \times C^\infty((-L,L)) \right)$ and to have value zero in a neighborhood of $x_1=L$. From the definition of $\mathcal{A}$, there exist smooth functions $\hat{\pmb{\xi}}$, characterizing the $KL_0$ displacement type. We recall that, in this particular case, $\check{\xi}_i=\vartheta=0$. Let us introduce the sequences $\hat{\mathbf{u}}^{\flat}_\e$ and $\e^k\check{\mathbf{u}}^{\flat}_\e$ defined as follows:

\begin{equation}\label{eq:recE}
	\begin{aligned}
		\hat{u}_{1\e}^\flat &\coloneqq \hat{\xi}_1(x_1, x_2) - x_3\partial_1 \hat{\xi}_3(x_1, x_2),\\
		\hat{u}_{2\e}^\flat &\coloneqq \hat{\xi}_2(x_1, x_2) - x_3\partial_2 \hat{\xi}_3(x_1, x_2),\\
		\hat{u}_{3\e}^\flat &\coloneqq \hat{\xi}_3(x_1, x_2) - \e^2 \frac{\nu}{1-\nu}\left[x_3\left(\partial_1 \hat{\xi}_1(x_1, x_2)+ \partial_2 \hat{\xi}_2(x_1, x_2)\right) - \frac{x_3^2}{2}\left(\partial^2_1 \hat{\xi}_3(x_1, x_2) + \partial^2_2 \hat{\xi}_3(x_1, x_2)\right) \right],\\[25pt]
		\e^k\check{u}_{1\e}^\flat&\coloneqq\e^k\hat{\xi}_1(x_1, \e^w x_2) - \e^{k+h-1} x_3\partial_1 \hat{\xi}_3(x_1, \e^w x_2),\\
		\e^k\check{u}_{2\e}^\flat&\coloneqq\e^{k+w}\hat{\xi}_2(x_1, \e^w x_2) - \e^{k+h-1+w} x_3\partial_2 \hat{\xi}_3(x_1, \e^w x_2),\\	
		\e^k\check{u}_{3\e}^\flat&\coloneqq\e^{k+h-1}\hat{\xi}_3(x_1, \e^w x_2) \begin{multlined}[t]- \frac{\nu}{1-\nu} \e^{k+2h} x_3\left(\partial_1 \hat{\xi}_1(x_1, \e^w x_2)+ \partial_2 \hat{\xi}_2(x_1, \e^w x_2)\right)\\ +\frac{\nu}{1-\nu} \e^{k+3h-1} \frac{x_3^2}{2}\left(\partial^2_1 \hat{\xi}_3(x_1, \e^w x_2) + \partial^2_2 \hat{\xi}_3(x_1, \e^w x_2)\right). \end{multlined}\\	
	\end{aligned}
\end{equation}
	It can be easily verified that the pair $(\hat{\mathbf{u}}^{\flat}_\e, \e^k\check{\mathbf{u}}^{\flat}_\e)$ satisfies the boundary conditions at $x_1 = L$, the junction conditions \eqref{eq:cond}, and verifies the following convergences
\begin{equation}\label{eq:babababa}
	\begin{aligned}
		\hat{\mathbf{E}}_\e \hat{\mathbf{u}}^{\flat}_\e &\to \hat{\pmb{Z}} & \text{ in } L^2(\hat{\Omega}, \mathbb{R}^{3\times 3}),\\
		\e^k \check{\mathbf{E}}_\e \check{\mathbf{u}}^{\flat}_\e & \to \pmb{0} & \text{ in } L^2(\check{\Omega}, \mathbb{R}^{3\times 3}),\\
		\e^{k+M} \left(\check{\mathbf{W}}_\e \check{\mathbf{u}}^{\flat}_\e\right)_{32} &\to 0 & \text{ in } L^2(\check{\Omega}), \\
		\hat{\mathbf{u}}^{\flat}_\e &\to \hat{\mathbf{u}} & \text{ in } W^{1,2}(\hat{\Omega}, \mathbb{R}^{3}), \\
		\e^k 
		\check{\mathbf{u}}^{\flat}_\e&\to \textbf{0} & \text{ in } W^{1,2}(\check{\Omega}, \mathbb{R}^{3}),
	\end{aligned}
\end{equation}
with $\hat{\pmb{Z}}$ defined as in \eqref{hatZr}.

The proof is concluded as in case {\it i)}
\end{proof}

\begin{myRemark}
As it can be noticed, the construction of the recovery sequence is quite cumbersome, and general construction rules do not exist. For this reason, we decided to provide the recovery sequence for two \textrm{extreme} cases (ii and iii) and for an \textrm{intermediate} one (i), see also Fig.\  \ref{fig:cases}. As a consequence, in the statement of Theorem \ref{gammatheorem} we have explicitly considered, in the part concerning the existence of a recovery sequence, only three out of the twenty-three possible cases. 
We stress the fact that the given \textit{Liminf inequality} proof is valid for every choice of the scaling parameters (in the admissible ranges), i.e. for all the twenty-three cases.\\
We did not check all the twenty-three cases, but in all the presented cases we checked we were able to construct a recovery sequence. We are therefore confident that Theorem \ref{gammatheorem} holds \textit{for all} the twenty-three cases, even if it is stated for only three.\\
\end{myRemark}

We can explicitly write the expression of the limit stored-energy functional.
Considering the case \textit{i)} of Theorem \ref{gammatheorem} for the sake of simplicity, we have:
\begin{equation*}
	\check{\mathcal{W}}(\check{\mathbf{u}}, \vartheta) = \frac{1}{2} \int_{-L}^{L} 
	\left[\begin{matrix}
		\textsf{E}A & -\textsf{E}S_2 & 0\\
		            & \textsf{E}J_2  & 0\\
		   \mathrm{sym} &  & \mu J_t
	\end{matrix}\right]\ \left(\begin{matrix}
		\partial_1 \hat{\xi}_{1}(x_1,0)\\ \partial^2_1\check{\xi}_{3}(x_1) \\ \partial_1 \vartheta(x_1)
	\end{matrix} \right) \cdot \left(\begin{matrix}
	\partial_1 \hat{\xi}_{1}(x_1,0)\\ \partial^2_1\check{\xi}_{3}(x_1) \\ \partial_1 \vartheta(x_1)
\end{matrix} \right) \diff x_1,
\end{equation*}
where 
$
A \coloneqq\int_{\check{\omega}} \diff x_2\diff x_3 = 2WH$ is the cross-sectional area of the stiffener,  $S_2 \coloneqq\int_{\check{\omega}} x_3 \diff x_2\diff x_3 = WH^2$ is the static moment with respect to the $x_2$ axis, 
$J_2 \coloneqq\int_{\check{\omega}} x_3^2 \diff x_2\diff x_3=\frac{2}{3}WH^3$ is the moment of inertia with respect to the $x_2$ axis, $J_t \coloneqq 4\int_{\check{\omega}} x_2^2 \diff x_2\diff x_3=\frac{8}{3}HW^3$ is the \textit{torsional} moment of inertia. 
Similarly, we have
\begin{equation*}
	\hat{\mathcal{W}}(\hat{\mathbf{u}}) = \frac{1}{2} \int_{\hat{\omega}}
	\left[\begin{matrix}
		\frac{T\textsf{E}}{1-\nu^2}\mathbf{K} & -\frac{T^2\textsf{E}}{2(1-\nu^2)}\mathbf{K}\\
		-\frac{T^2\textsf{E}}{2(1-\nu^2)}\mathbf{K}^\mathrm{T} & \frac{T^3\textsf{E}}{3(1-\nu^2)}\mathbf{K}
	\end{matrix}\right]\ \left(\begin{matrix}
		\partial_1\hat{\xi}_{1}\\ \partial_2\hat{\xi}_{2}\\ \frac{1}{2}\left[ \partial_2\hat{\xi}_{1}+\partial_1\hat{\xi}_{2}\right] \\ \partial^2_1\hat{\xi}_{3}\\ \partial^2_2\hat{\xi}_{3}\\ \partial_1\partial_2\hat{\xi}_{3}
	\end{matrix} \right) \cdot \left(\begin{matrix}
	\partial_1\hat{\xi}_{1}\\ \partial_2\hat{\xi}_{2}\\ \frac{1}{2}\left[ \partial_2\hat{\xi}_{1}+\partial_1\hat{\xi}_{2}\right] \\ \partial^2_1\hat{\xi}_{3}\\ \partial^2_2\hat{\xi}_{3}\\ \partial_1\partial_2\hat{\xi}_{3}
\end{matrix} \right)
	\diff x_1 \diff x_2,
\end{equation*}
where we have posed
\begin{equation*}
		\mathbf{K} \coloneqq 
		\left[\begin{matrix}
			1 & \nu &0\\
			&   1 & 0\\
			\mathrm{sym} & & 2(1-\nu)
		\end{matrix}\right].
\end{equation*}

\section{Strong Convergence of Minima and Minimizers}\label{sec:congergence}
So far, we just considered the stored energy functional. However, the equilibrium problem is ruled by the total energy $\widetilde{\mathcal{F}}_\e$, which is the sum of the stored energy $\widetilde{\mathcal{W}}_\e$ minus the work done by external loads $\widetilde{\mathcal{L}}_\e$
\begin{equation*}
	\widetilde{\mathcal{F}}_\e(\mathbf{u}) \coloneqq \widetilde{\mathcal{W}}_\e(\mathbf{u}) -\widetilde{\mathcal{L}}_\e(\mathbf{u}),
\end{equation*}
where the work of external loads  is assumed to be
\begin{equation*}
	\widetilde{\mathcal{L}}_{\e}(\mathbf{u})\coloneqq\int_{\Omega_{\e}} \mathbf{b} \cdot \mathbf{u} \diff x = \int_{\hat{\Omega}_{\e}} \chi_\e(x) \mathbf{b} \cdot \mathbf{u} \diff x + \int_{\check{\Omega}_{\e}} \chi_\e(x) \mathbf{b} \cdot \mathbf{u} \diff x,
\end{equation*}
and where $\mathbf{b}$ belongs to $L_2(\Omega_\e)$.
After the scaling of Section \ref{sec:scaledproblem}, we have
\begin{equation*}
	\begin{aligned}
		 \mathcal{L}_{\e}(\hat{\mathbf{u}}_{\e}, \check{\mathbf{u}}_{\e})\coloneqq  \frac{1}{\e} \widetilde{\mathcal{L}}_{\e}(\mathbf{u}_{\e}) &= \int_{\hat{\Omega}} \hat{\chi}_\e \hat{\mathbf{b}}_\e \cdot \hat{\mathbf{u}}_\e \diff x + \int_{\check{\Omega}} \check{\chi}_\e \check{\mathbf{b}}_\e \cdot \e^k\check{\mathbf{u}}_\e \diff x\\
		&=: \hat{\mathcal{L}}_{\e}(\hat{\mathbf{u}}_{\e}) + \check{\mathcal{L}}_{\e}(\e^k \check{\mathbf{u}}_{\e}),
	\end{aligned}
\end{equation*}
where we have posed
\begin{equation*}
	\begin{aligned}
		\hat{\mathbf{b}}_\e &\coloneqq  \hat{\mathbf{Q}}^{-\mathrm{T}}_\e \mathbf{b}\circ \hat{\mathbf{q}}_\e,  &\check{\mathbf{b}}_\e &\coloneqq  \e^k\check{\mathbf{Q}}^{-\mathrm{T}}_\e \mathbf{b}\circ \check{\mathbf{q}}_\e.
	\end{aligned}
\end{equation*}
In particular, we consider loads of the form
\begin{equation*}
	\begin{aligned}
		\hat{\mathbf{b}}_\e &= \left( \hat{b}_1(x), \hat{b}_2(x), \hat{b}_3(x)\right),\\
		\check{\mathbf{b}}_\e &= \left(\check{b}_1(x), \check{b}_2(x) - \e^{-m} \frac{m(x_1)}{I_G}(x_3 - x_3(G)),  \check{b}_3(x) + \e^{-m} \frac{m(x_1)}{I_G}(x_2 - x_2(G))
		\right) .
	\end{aligned}
\end{equation*}
where $\hat{b}_i \in L^2(\hat{\Omega})$, $\check{b}_i \in L^2(\check{\Omega})$ and $m(x_1)\in L^2((-L,L))$. 

It is easy to see that the external loads contributions continuously converge in the sense of the convergence used in Theorem \ref{gammatheorem} to the limit functionals
\begin{equation*}
	\begin{aligned}
		\hat{\mathcal{L}}_{\e}(\hat{\mathbf{u}}_{\e}) \to \hat{\mathcal{L}}(\hat{\mathbf{u}}) &= \int_{\hat{\Omega}} \hat{b}_i \hat{u}_i \diff x \\
		\check{\mathcal{L}}_{\e}(\e^k \check{\mathbf{u}}_{\e}) \to \check{\mathcal{L}}(\check{\mathbf{u}}, \vartheta) &= \int_{\check{\Omega}} \check{b}_i \check{u}_i \diff x + \int_{-L}^L m\vartheta \diff x_1.\\
	\end{aligned}
\end{equation*}

In the next Theorem, we show that the external loads do not impact on our $\Gamma$-limit result. This is the reason why we focused only on the stored energy in the previous part of the paper.
\begin{myTheorem}\label{th:71}
	As $\e\downarrow 0$, the sequence of functionals $\mathcal{F}_\e (\hat{\mathbf{u}}_\e, \e^k \check{\mathbf{u}}_\e)\coloneqq \mathcal{W}_\e(\hat{\mathbf{u}}_\e, \e^k \check{\mathbf{u}}_\e) - \mathcal{L}_\e(\hat{\mathbf{u}}_\e, \e^k \check{\mathbf{u}}_\e) $ $\Gamma$-converges to the limit functional $\mathcal{F}(\hat{\mathbf{u}}, \check{\mathbf{u}}, \vartheta)\coloneqq \mathcal{W}(\hat{\mathbf{u}}, \check{\mathbf{u}}, \vartheta) - \mathcal{L}(\hat{\mathbf{u}}, \check{\mathbf{u}}, \vartheta)$ in the sense specified in Theorem \ref{gammatheorem}.
\end{myTheorem}
\begin{proof}
	The proof follows from the well-known stability of $\Gamma$-convergence with respect to continuous, real-valued perturbations (see \cite[Proposition $6.20$]{Maso1993}).
	%\QEDB
\end{proof}

We conclude the paper by showing that the sequence of minima and minimizers from the sequence of three-dimensional total energies converges to the unique solution of the variational $\Gamma$-limit problem. From a mechanical point of view, it can be interpreted as follows: the equilibrium configurations of the sequence of three-dimensional problems converge towards the equilibrium configuration provided by the $\Gamma$-limit functional minimization. Moreover, we show that the convergence is actually strong.
\begin{myTheorem}
	Suppose Theorems \ref{gammatheorem} and \ref{th:71} hold.\\
	As $\e\downarrow 0$, the sequence of three-dimensional minimization problems for the functional
	$\mathcal{F}_\e(\hat{\textbf{u}}_\e, \e^k\check{\textbf{u}}_\e)\coloneqq\mathcal{W}_\e(\hat{\textbf{u}}_\e, \e^k\check{\textbf{u}}_\e)-\mathcal{L}_\e(\hat{\textbf{u}}_\e, \e^k\check{\textbf{u}}_\e)$
		\begin{equation}\label{eq:problim3d}
		\begin{multlined}
			\min\limits_{(\hat{\textbf{u}}_\e, \e^k\check{\textbf{u}}_\e)\in \mathcal{A}_\e} \mathcal{F}_\e(\hat{\textbf{u}}_\e, \e^k\check{\textbf{u}}_\e)
		\end{multlined}
	\end{equation}
	 has a unique solution for each term in the sequence. The solution is denoted by $(\hat{\mathbf{u}}^{\#}_{\e}, \e^k \check{\mathbf{u}}^{\#}_{\e})$.\\
	Similarly, the minimization problem for the $\Gamma$-limit functional $\mathcal{F}(\hat{\pmb{u}}, \check{\pmb{u}}, \vartheta)\coloneqq\mathcal{W}(\hat{\pmb{u}}, \check{\pmb{u}}, \vartheta)-\mathcal{L}(\hat{\pmb{u}}, \check{\pmb{u}}, \vartheta)$
	\begin{equation}\label{eq:problim}
		\begin{multlined}
			\min\limits_{(\hat{\mathbf{u}}, \check{\mathbf{u}}, \vartheta)\in \mathcal{A}} \mathcal{F}(\hat{\pmb{u}}, \check{\pmb{u}}, \vartheta)
		\end{multlined}
	\end{equation}
	admits a unique solution denoted by $(\hat{\pmb{u}}^{\#}, \check{\pmb{u}}^{\#}, \vartheta^{\#})$. Moreover, we have that
	\begin{enumerate}
		\item $\hat{\mathbf{u}}^{\#}_{\e} \to \hat{\mathbf{u}}^{\#} \text{ in } W^{1,2}(\hat{\Omega},  \mathbb{R}^3)$;\label{pt1}
		\item $ \e^k \check{\mathbf{u}}^{\#}_{\e} \to \check{\mathbf{u}}^{\#} \text{ in } W^{1,2}(\check{\Omega},  \mathbb{R}^3)$;\label{pt2}
		\item $\e^{k+M} \left(\check{\mathbf{W}}_\e \check{\mathbf{u}}^{\#}_{\e}\right)_{32} \to \vartheta^{\#} \text{ in } L^2(\check{\Omega})$;\label{pt32}
		\item $\mathcal{F}_\e(\hat{\mathbf{u}}^{\#}_{\e}, \e^k \check{\mathbf{u}}^{\#}_{\e}) \text{ converges to } \mathcal{F}(\hat{\mathbf{u}}^{\#}, \check{\mathbf{u}}^{\#}, \vartheta^{\#})$.\label{pt3}
	\end{enumerate}
\end{myTheorem}
\begin{proof}
	
	The existence of a solution for problems \eqref{eq:problim3d}, \eqref{eq:problim}
	can be proved through the Direct Method of the Calculus of Variations; the uniqueness follows from the strict convexity of the functionals $\mathcal{F}_\e$ and $\mathcal{F}$.
	
	From Theorem \ref{th:71}, Propositions 6.8 and 8.16 (lower-semicontinuity of sequential $\Gamma$-limits), Theorem 7.8 (coercivity of $\Gamma$-limits) and Corollary 7.24 (convergence of minima and minimizer) of \cite{Maso1993}, it follows that the weak convergence counterpart of points \ref{pt1}, \ref{pt2}, \ref{pt32} is satisfied, and that point \ref{pt3} is also proved.
	To show that the convergence is actually strong, we adapt some arguments proposed in \cite{Freddi2004, Freddi2007}.

	Let us denote by $\mathbf{a}_\e$ the approximate minimizer of problem (\ref{eq:problim}), defined as the recovery sequence(s) appearing in Theorem \ref{gammatheorem}, but with $(\hat{\pmb{\xi}}, \check{\pmb{\xi}}, \vartheta)$ replaced by $(\hat{\pmb{\xi}}^{\#}, \check{\pmb{\xi}}^{\#}, \vartheta^{\#})$,  related to the pair $(\hat{\mathbf{u}}^{\#}, \check{\mathbf{u}}^{\#})$.  
	By part {\it (b)} of Theorem  \ref{gammatheorem} and by Theorem \ref{th:71} we have
	\begin{equation*}
		\begin{aligned}
			\lim\limits_{\e\downarrow 0} \mathcal{F}_\e(\hat{\mathbf{a}}_\e, \e^k\check{\mathbf{a}}_\e) &= \mathcal{F}(\hat{\pmb{\xi}}^{\#}, \check{\pmb{\xi}}^{\#}, \vartheta^{\#}),
			&
			\lim\limits_{\e\downarrow 0} \mathcal{L}_\e(\hat{\mathbf{a}}_\e, \e^k \check{\mathbf{a}}_\e) &= \mathcal{L}(\hat{\pmb{\xi}}^{\#}, \check{\pmb{\xi}}^{\#}, \vartheta^{\#}).
		\end{aligned}
	\end{equation*}
	In particular,  
	\begin{equation}\label{eq:pippo}
		\begin{aligned}		
			\liminf\limits_{\e\downarrow 0} \left(\hat{\mathcal{F}}_\e(\hat{\mathbf{u}}^{\#}_{\e}) - \hat{\mathcal{F}}_\e(\hat{\mathbf{a}}_{\e})\right)&\le 0,
			&
			\liminf\limits_{\e\downarrow 0} \left(\check{\mathcal{F}}_\e(\e^k\check{\mathbf{u}}^{\#}_{\e}) - \check{\mathcal{F}}_\e(\e^k\check{\mathbf{a}}_{\e})\right)&\le 0,\\
			\lim\limits_{\e\downarrow 0} \hat{\mathcal{L}}_\e(\hat{\mathbf{u}}^{\#}_{\e} - \hat{\mathbf{a}}_{\e})&=0			
			&
			\lim\limits_{\e\downarrow 0} \check{\mathcal{L}}_\e(\e^k\check{\mathbf{u}}^{\#}_{\e} - \e^k\check{\mathbf{a}}_{\e})&= 0.
		\end{aligned}
	\end{equation}
	As a preliminary observation, quadratic forms (\ref{f}) satisfy the identity
	\begin{equation*}
		f(\mathbf{U}) = f(\mathbf{A}) + \mathbb{C}[\mathbf{A}]\cdot(\mathbf{U}-\mathbf{A}) + f(\mathbf{U}-\mathbf{A})
	\end{equation*}
	for every $\mathbf{A}$, $\mathbf{U} \in  \mathbb{R}^{3 \times 3}$. By the coercivity condition (\ref{coercivity}), we obtain the following inequality: 
	\begin{equation*}
		f(\mathbf{U}) \geq f(\mathbf{A}) + \mathbb{C}[\mathbf{A}]\cdot(\mathbf{U}-\mathbf{A}) + \mu |\mathbf{U}-\mathbf{A}|^2.
	\end{equation*}
	Then, we have
	\begin{equation}\label{eqqq}
		\begin{aligned}
			\hat{\mathcal{F}}_\e(\hat{\mathbf{u}}^{\#}_{\e}) - \hat{\mathcal{F}}_\e(\hat{\mathbf{a}}_{\e}) &\geq \int_{\hat{\Omega}}\frac{1}{2}\mathbb{C}\left[\hat{\mathbf{E}}_\e \hat{\mathbf{a}}_\e\right]\cdot \hat{\mathbf{E}}_\e (\hat{\mathbf{u}}^{\#}_{\e}-\hat{\mathbf{a}}_\e) \diff x + \frac{\mu}{2}  \norm{\hat{\mathbf{E}}_\e \hat{\mathbf{u}}^{\#}_{\e} -\hat{\mathbf{E}}_\e \hat{\mathbf{a}}_{\e} }^2_{L^2(\hat{\Omega}, \mathbb{R}^{3\times 3})}-\hat{\mathcal{L}}_\e(\hat{\mathbf{u}}^{\#}_{\e}-\hat{\mathbf{a}}_\e),\\
			\check{\mathcal{F}}_\e(\e^k\check{\mathbf{u}}^{\#}_{\e}) - \check{\mathcal{F}}_\e(\e^k\check{\mathbf{a}}_{\e}) &\geq \int_{\check{\Omega}}\frac{1}{2}\mathbb{C}\left[\e^k\check{\mathbf{E}}_\e \check{\mathbf{a}}_\e\right]\cdot \e^k \check{\mathbf{E}}_\e (\check{\mathbf{u}}^{\#}_{\e}-\check{\mathbf{a}}_\e) \diff x + \frac{\mu}{2}  \norm{\e^k\check{\mathbf{E}}_\e \check{\mathbf{u}}^{\#}_{\e} -\e^k\check{\mathbf{E}}_\e \check{\mathbf{a}}_{\e} }^2_{L^2(\check{\Omega}, \mathbb{R}^{3\times 3})}-\check{\mathcal{L}}_\e(\e^k\check{\mathbf{u}}^{\#}_{\e}-\e^k\check{\mathbf{a}}_\e),
		\end{aligned}
	\end{equation}
since, $\hat{\chi}_\e,\,\check{\chi}_\e \geq \frac{1}{2}$.
	For brevity sake, we introduce the following notation (to be specialized for the plate and the stiffener with the usual symbols $\hat{\dot}$ and $\check{\dot}$ ):
	\begin{equation*}
		\begin{aligned}
			\mathbf{A}_\e &\coloneqq \mathbf{E}_\e \mathbf{a}_\e,  &\mathbf{U}_\e&\coloneqq\mathbf{E}_\e \mathbf{u}^{\#}_{\e}, &\pmb{\Delta}_\e &\coloneqq \mathbf{U}_\e-\mathbf{A}_\e,
		\end{aligned}
	\end{equation*}
	so that (\ref{eqqq}) rewrites
	\begin{equation*}
		\begin{aligned}
			\hat{\mathcal{F}}_\e(\hat{\mathbf{u}}^{\#}_{\e}) - \hat{\mathcal{F}}_\e(\hat{\mathbf{a}}_{\e}) &\geq \int_{\hat{\Omega}}\frac{1}{2}\mathbb{C}\left[\hat{\mathbf{A}}_\e\right]\cdot \hat{\pmb{\Delta}}_\e \diff x + \frac{\mu}{2}  \norm{\hat{\pmb{\Delta}}_\e} ^2_{L^2(\hat{\Omega}, \mathbb{R}^{3\times 3})}-\hat{\mathcal{L}}_\e(\hat{\mathbf{u}}^{\#}_{\e}-\hat{\mathbf{a}}_\e),\\
			\check{\mathcal{F}}_\e(\e^k\check{\mathbf{u}}^{\#}_{\e}) - \check{\mathcal{F}}_\e(\e^k\check{\mathbf{a}}_{\e}) &\geq \int_{\check{\Omega}}\frac{1}{2}\mathbb{C}\left[\e^k\check{\mathbf{A}}_\e\right]\cdot \e^k\check{\pmb{\Delta}}_\e \diff x + \frac{\mu}{2}   \norm{\e^k\check{\pmb{\Delta}}_\e} ^2_{L^2(\check{\Omega}, \mathbb{R}^{3\times 3})}-\check{\mathcal{L}}_\e(\e^k\check{\mathbf{u}}^{\#}_{\e}-\e^k\check{\mathbf{a}}_\e).
		\end{aligned}
	\end{equation*}
	To start, we prove that the first integral appearing in the right hand side of both of \eqref{eqqq} tends to zero in the limit of $\e\downarrow 0$.
The integrands of these two integrals rewrite as 
	\begin{equation}\label{eq:integrands}
		\begin{aligned}
			2\mu\left(\hat{A}_{ij\e}\hat{\Delta}_{ij\e}\right) &+ \lambda\left(\hat{A}_{ii\e}\hat{\Delta}_{jj\e}\right),
			& \e^{2k} [2\mu\left(\check{A}_{ij\e}\check{\Delta}_{ij\e}\right) &+ \lambda\left(\check{A}_{ii\e}\check{\Delta}_{jj\e}\right)].
		\end{aligned}
	\end{equation}
	It follows by part {\it (b)} of Theorem \ref{gammatheorem} that	
	\begin{equation}\label{eq:37}
		\begin{aligned}
			\hat{\mathbf{A}}_\e &\to \hat{\pmb{Z}} &\text{ in } L^2(\hat{\Omega},  \mathbb{R}^{3\times 3}),\\
			\e^k\check{\mathbf{A}}_\e &\to \check{\pmb{Z}} &\text{ in } L^2(\check{\Omega}, \mathbb{R}^{3\times 3}),
		\end{aligned}
	\end{equation}
with $\hat{\pmb{Z}}$ and $\check{\pmb{Z}}$ defined as in \eqref{eq:energybeam}.

By Lemma \ref{compattezzapiastra}, Lemma \ref{compattezzatrave} and \eqref{eq:37} we have that $\hat{\pmb{\Delta}}_\e$ and $\e^k\check{\pmb{\Delta}}_\e$ are bounded in $L^2(\hat{\Omega}, \mathbb{R}^{3\times 3})$ and $L^2(\check{\Omega}, \mathbb{R}^{3\times 3})$, respectively.
	Thus, from (\ref{eq:37}) and the structure of $\hat{\pmb{Z}}$ and $\check{\pmb{Z}}$, we deduce  that 
	\begin{equation*}
		\begin{aligned}
			&\lim\limits_{\e\downarrow 0} \int_{\hat{\Omega}} \hat{A}_{\alpha 3 \e}\hat{\Delta}_{\alpha 3 \e} \diff x =0,\\
			&\begin{cases}
				\lim\limits_{\e\downarrow 0} \int_{\check{\Omega}} \e^{2k}\check{A}_{23 \e}\check{\Delta}_{23 \e} \diff x =0,  & \text{ if } M=w=h,\\
				\lim\limits_{\e\downarrow 0} \int_{\check{\Omega}} \e^{2k}\check{A}_{23 \e}\check{\Delta}_{23 \e} \diff x = \lim\limits_{\e\downarrow 0} \int_{\check{\Omega}} \e^{2k}\check{A}_{12 \e}\check{\Delta}_{12 \e} \diff x = 0,  & \text{ if } M=w\neq h,\\
				\lim\limits_{\e\downarrow 0} \int_{\check{\Omega}} \e^{2k}\check{A}_{23 \e}\check{\Delta}_{23 \e} \diff x = \lim\limits_{\e\downarrow 0} \int_{\check{\Omega}} \e^{2k}\check{A}_{13 \e}\check{\Delta}_{13 \e} \diff x = 0,  & \text{ if } M=h\neq w.
			\end{cases}
		\end{aligned}
	\end{equation*}
	From \eqref{Hhat}, \eqref{eq:Hcheck}, and (\ref{eq:37}), it follows that $\hat{\Delta}_{\alpha\beta}\weak 0$ in $L^2(\hat{\Omega})$ and that $\e^k\check{\Delta}_{11}\weak 0$ in $L^2(\check{\Omega})$, and therefore
	\begin{equation*}
		\begin{aligned}
			\lim\limits_{\e\downarrow 0} \int_{\hat{\Omega}} \hat{A}_{ij\e}\hat{\Delta}_{\alpha\beta \e} \diff x &=0,
			&\lim\limits_{\e\downarrow 0} \int_{\check{\Omega}} \e^{2k}\check{A}_{ij\e}\check{\Delta}_{11\e} \diff x &=0.
		\end{aligned}
	\end{equation*}
	From Lemma \ref{otherscond}, it also follows that, up to subsequences,
	\begin{equation*}
		\begin{aligned}
			\e^k\check{U}_{13 \e} &\weak 
			\begin{cases}
				\frac{1}{2}\left(\partial_3\Phi^{\#} + x_2\right)\partial_1\vartheta^{\#}  & \textrm{ if } M=w=h,\\
				x_2 \partial_1\vartheta^{\#}+\eta_{13}^\# & \textrm{ if } M=w\neq h,\\
				0 & \textrm{ otherwise },
			\end{cases}\\
			\e^k\check{U}_{12 \e} &\weak  
			\begin{cases}
				\frac{1}{2}\left(\partial_2\Phi^{\#} - (x_3 - \frac{H}{2})\right)\partial_1\vartheta^{\#} & \textrm{ if } M=w=h,\\
				-(x_3 - \frac{H}{2}) \partial_1\vartheta^{\#}+\eta_{12}^\# & \textrm{ if } M=h\neq w,\\
				0 & \textrm{ otherwise },
			\end{cases}
		\end{aligned}
	\end{equation*}
	in $L^2(\check{\Omega})$, where $\Phi^{\#}$, $\eta_{12}^{\#}$ and $\eta_{13}^{\#}$ are specified in the Lemma. 	By (\ref{eq:37}), 
	\begin{equation*}
		\begin{aligned}
			\e^k\check{A}_{13 \e}&\to 
			\begin{cases}
				\frac{1}{2}\left(\partial_3\Phi^{\#} + x_2\right)\partial_1\vartheta^{\#}  & \textrm{ if } M=w=h,\\
				x_2 \partial_1\vartheta^{\#} & \textrm{ if } M=w\neq h,\\
				0 & \textrm{ otherwise },
			\end{cases}\\
			\e^k\check{A}_{12 \e}&\to 
			\begin{cases}
				\frac{1}{2}\left(\partial_2\Phi^{\#} - (x_3 - \frac{H}{2})\right)\partial_1\vartheta^{\#} & \textrm{ if } M=w=h,\\
				-(x_3 - \frac{H}{2}) \partial_1\vartheta^{\#} & \textrm{ if } M=h\neq w,\\
				0 & \textrm{ otherwise },
			\end{cases}
		\end{aligned}
	\end{equation*}
 in $L^2(\check{\Omega})$. In any case,
	\begin{equation*}
		\lim\limits_{\e\downarrow 0} \int_{\check{\Omega}} \e^{2k}\check{A}_{1a\e}\check{\Delta}_{1a\e} \diff x = 0.
	\end{equation*}
	Let $\hat{\Delta}_{33}$ be the limit in the weak $L^2(\hat{\Omega})$ topology of $\hat{\Delta}_{33\e}$, and $\check{\Delta}_{aa}$ be the limit in the weak $L^2(\check{\Omega})$ topology of $\e^k\check{\Delta}_{aa\e}$.
	Summarizing, we have
	\begin{equation*}
		\begin{split}
			\lim\limits_{\e\downarrow 0} \int_{\hat{\Omega}}\mathbb{C}\hat{\mathbf{A}}_\e \cdot \hat{\Delta}_\e \diff x &=  \lim\limits_{\e\downarrow 0} \int_{\hat{\Omega}} 2\mu\hat{A}_{33\e} \hat{\Delta}_{33\e} + \lambda \hat{A}_{ii\e} \hat{\Delta}_{33\e} \diff x\\ 
			&= \int_{\hat{\Omega}} \hat{\Delta}_{33}(\partial_1\hat{u}_{1\#} + \partial_2\hat{u}_{2\#})\frac{\lambda(1-2\nu)-2\mu\nu}{1-\nu}\diff x = 0,\\
			\lim\limits_{\e\downarrow 0} \int_{\check{\Omega}}\e^{2k}\mathbb{C}\check{\mathbf{A}}_\e \cdot\check{\Delta}_\e \diff x &=  \lim\limits_{\e\downarrow 0} \int_{\check{\Omega}} \e^{2k} \left[ 2\mu\check{A}_{aa\e} \check{\Delta}_{aa\e} + \lambda \check{A}_{ii\e} \check{\Delta}_{bb\e} \right] \diff x\\ 
			&= \int_{\check{\Omega}} \check{\Delta}_{aa}\partial_1\check{u}_{\#1}(\lambda-2\nu(\mu+\lambda))\diff x = 0,
		\end{split}
	\end{equation*}
	because of the relationships between elastic moduli ($\textsf{E}$, $\nu$) and Lamé constants ($\lambda$, $\mu$) nullify identically the integrand functions. 
	Hence, considering  (\ref{eq:pippo}) and (\ref{eqqq}), we have shown that
	\begin{equation}\label{deltazero}
		\begin{aligned}
			\norm{\hat{\mathbf{E}}_\e \hat{\mathbf{u}}^{\#}_{\e} -\hat{\mathbf{E}}_\e \hat{\mathbf{a}}_{\e} }^2_{L^2(\hat{\Omega}, \mathbb{R}^{3\times 3})}&\to 0,
			&\norm{\e^k \check{\mathbf{E}}_\e \check{\mathbf{u}}^{\#}_{\e} -\e^k\check{\mathbf{E}}_\e \check{\mathbf{a}}_{\e} }^2_{L^2(\check{\Omega}, \mathbb{R}^{3\times 3})}&\to 0.
		\end{aligned}
	\end{equation}
	By applying Korn inequality, we also have that 
	\begin{equation*}
		\begin{aligned}
			\norm{\hat{\mathbf{u}}^{\#}_{\e}-\hat{\mathbf{a}}_\e}_{W^{1,2}(\hat{\Omega}, \mathbb{R}^{3})}&\to 0,
			&\norm{\e^k\check{\mathbf{u}}^{\#}_{\e}-\e^k\check{\mathbf{a}}_\e}_{W^{1,2}(\check{\Omega}, \mathbb{R}^{3})}&\to 0.
		\end{aligned}
	\end{equation*}
	From which points \ref{pt1} and \ref{pt2} follow.\\
	Furthermore, by (\ref{deltazero}) and Lemma \ref{lemmatheta}, we have 
	\begin{equation*}
		\lim\limits_{\e\downarrow 0}\norm{\e^{k+M} \check{\mathbf{H}}_\e(\check{\mathbf{u}}^{\#}_{\e}-\check{\mathbf{a}}_\e)}^2_{L^2(\check{\Omega})}=0,
	\end{equation*}
	that implies that
	\begin{equation*}
		\e^{k+M} \left(\check{\mathbf{W}}_\e\check{\mathbf{u}}^{\#}_{\e}\right)_{32} = \e^{k+M}\left(\check{\mathbf{H}}_\e(\check{\mathbf{u}}^{\#}_{\e}-\check{\mathbf{a}}_\e)\right)_{32}-\e^{k+M}\left(\check{\mathbf{H}}_\e(\check{\mathbf{u}}^{\#}_{\e}-\check{\mathbf{a}}_\e)\right)_{23} + \e^{k+M}\left(\check{\mathbf{W}}_\e\check{\mathbf{a}}_\e\right)_{32}\to \vartheta^{\#}
	\end{equation*}
	in $L^2(\check{\Omega})$. Hence, also point \ref{pt32} is proven.
\end{proof}

\begin{myRemark}
Being confident that the $\Gamma$-convergence result of Theorem \ref{gammatheorem} can be extended to all the twenty-three cases, Theorem \ref{th:71} actually assesses the strong convergence of minima and minimizers for all such cases.
\end{myRemark}

\section*{Acknowledgements}
MPS gratefully acknowledges the support of
project PARSIFAL ("PrandtlPlane ARchitecture for the Sustainable Improvement of
Future AirpLanes"), which has been funded by the European Union under the
Horizon 2020 Research and Innovation Program (Grant Agreement n.723149).

\bibliographystyle{ieeetr}

\end{document}